\documentclass[
     reprint,
    aps,
    prd, 
    superscriptaddress,
    showpacs,
    nofootinbib,
    amsmath,
    amssymb,
    longbibliography 
]{revtex4-2}
\usepackage{slashed}
\usepackage{graphicx}
\usepackage{dcolumn}
\usepackage{bm}
\usepackage{hyperref}
\usepackage{color}
\usepackage{orcidlink}
\usepackage{filecontents}
\usepackage{float}
\usepackage{booktabs}
\usepackage{mathtools}
\usepackage{tikz}

\usetikzlibrary{arrows.meta, positioning, shapes.geometric, calc, fit, backgrounds}

\usepackage{xcolor}
\definecolor{darkblue}{rgb}{0.0, 0.0, 0.55}
\definecolor{darkred}{rgb}{0.55, 0.0, 0.0}

\hypersetup{
    colorlinks=true,
    linkcolor=darkblue, 
    citecolor=darkred,  
    urlcolor=darkblue   
}

\begin{document}

\title{Probing Proton Structure via Physics-Guided Neural Networks in Holographic QCD}

\author{Wei Kou\orcidlink{0000-0002-4152-2150}}
\email{kouwei@impcas.ac.cn}
\affiliation{Institute of Modern Physics, Chinese Academy of Sciences, Lanzhou 730000, Gansu Province, China}
\affiliation{Southern Center for Nuclear Science Theory (SCNT), Institute of Modern Physics, Chinese Academy of Sciences, Huizhou 516000, Guangdong Province, China}
\affiliation{School of Nuclear Science and Technology, University of Chinese Academy of Sciences, Beijing 100049, China}
\affiliation{State Key Laboratory of Heavy Ion Science and Technology, Institute of Modern Physics, Chinese Academy of Sciences, Lanzhou 730000, Gansu Province, China}

\author{Xurong Chen}
\email{xchen@impcas.ac.cn}
\affiliation{Institute of Modern Physics, Chinese Academy of Sciences, Lanzhou 730000, Gansu Province, China}
\affiliation{Southern Center for Nuclear Science Theory (SCNT), Institute of Modern Physics, Chinese Academy of Sciences, Huizhou 516000, Guangdong Province, China}
\affiliation{School of Nuclear Science and Technology, University of Chinese Academy of Sciences, Beijing 100049, China}
\affiliation{State Key Laboratory of Heavy Ion Science and Technology, Institute of Modern Physics, Chinese Academy of Sciences, Lanzhou 730000, Gansu Province, China}

\begin{abstract}
	Describing the proton structure function $F_2(x,Q^2)$ across the
nonperturbative and transition regimes of quantum chromodynamics remains a
significant theoretical challenge. We construct a physics-guided
phenomenological representation that combines fermion-resonance and
Pomeron-like AdS/QCD basis functions with a neural parametrization. The
fermion-resonance sector is constrained by the AdS eigenvalue problem and
the physical proton mass, while the neural parametrization describes the
effective kinematic mixing, the Pomeron normalization, and a regularized
log-space correction. For 187 selected SLAC measurements, a fixed
training--test partition containing 153 and 34 points, respectively,
yields the point-normalized diagnostics
$\chi^2/N_{\rm train}=0.542$,
$\chi^2/N_{\rm test}=1.990$, and
$\chi^2/N_{\rm all}=0.805$. Conditional on the chosen holographic basis,
the fitted channel fractions show a continuous transition from
Pomeron-like dominance at the lower end of the measured $x$ range to
fermion-resonance dominance at larger $x$. The midpoint of the mixing
function occurs at $x\simeq0.187$, whereas equality of the final channel
fractions occurs at $x\simeq0.226$. These two definitions identify a
transition region of approximately $x\simeq0.19$--$0.23$, rather than a
unique physical boundary. Although $\alpha_0\simeq1.08$ is compatible with
the data, fixed-$\alpha_0$ refits show that it is not separately
identifiable from the flexible Pomeron normalization. The $x$-dependent
holographic coordinates and associated effective excited scales should
therefore be interpreted as basis-dependent effective quantities rather
than direct measurements of an $x$-dependent proton spectrum. A comparison
with a data-only neural baseline does not establish a predictive advantage
for the holographic basis on the present dataset. The resulting framework
thus provides a basis-conditioned phenomenological decomposition of the
data, rather than a model-independent extraction of the underlying proton
dynamics.
\end{abstract}

\maketitle

\section{Introduction}
\label{sec:intro}

Deep inelastic scattering (DIS) is a principal probe of the internal structure of the nucleon \cite{Bloom:1969kc,Breidenbach:1969kd,Whitlow:1990gk,Whitlow:1991uw}. Fixed-target measurements performed at the Stanford Linear Accelerator Center (SLAC) provide important information on the proton structure function $F_2(x,Q^2)$ across the valence-dominated region and the transition between resonance-like and deep-inelastic behavior \cite{Whitlow:1990gk,Whitlow:1991uw}. In the low-to-moderate-$Q^2$ and moderate-to-large-$x$ regime, however, a leading-twist perturbative description alone is generally insufficient. Target-mass effects, higher-twist contributions, confinement dynamics, and remnants of hadronic resonance structure may all become relevant \cite{Brodsky:1989pv,Radyushkin:1983mj,Jaffe:1982pm,Ellis:1982wd,
	Roberts:1994dr,Alkofer:2000wg}. A uniformly controlled first-principles calculation of $F_2$ throughout this kinematic region therefore remains challenging and motivates the use of nonperturbative effective descriptions.

Gauge/gravity duality \cite{Maldacena:1997re,Gubser:1998bc,Witten:1998qj} provides one such route to modeling strongly coupled hadronic dynamics. In particular, soft-wall AdS/QCD models reproduce approximately linear hadronic mass trajectories and supply normalizable bulk wave functions for hadronic states \cite{Karch:2006pv,Andreev:2006ct,Gutsche:2011vb}. Within holographic descriptions of DIS, two complementary constructions have been used. At larger $x$, transitions of a bulk fermion into excited states provide an $s$-channel description of resonance-like contributions \cite{Polchinski:2002jw,Braga:2011wa,FolcoCapossoli:2020pks,
	Nascimento:2025ryr,Deng:2025fpq,Deng:2026gik}. Toward smaller $x$, $t$-channel holographic Pomeron exchange provides a description of the smooth high-energy background \cite{BallonBayona:2007rs,FolcoCapossoli:2015hub,Watanabe:2019zny}. These constructions describe different aspects and kinematic limits of holographic DIS. In the present work, they are adopted as two prescribed basis components within a common phenomenological representation rather than derived as sectors of a new unified dynamical equation. Moreover, within the SLAC kinematic region considered here, the Pomeron-like term should be understood as an effective smooth-background contribution, not as a precision determination of small-$x$ Regge dynamics.

Machine-learning parametrizations have become useful in high-energy and nuclear physics \cite{Larkoski:2017jix,Guest:2018yhq,Radovic:2018dip,
	Albertsson:2018maf,Carleo:2019ptp,Boehnlein:2021eym,
	Aarts:2025gyp}, including global analyses of parton distribution functions \cite{DelDebbio:2004xtd,Ball:2012cx,NNPDF:2014otw,
	NNPDF:2017mvq,Carrazza:2019mzf,NNPDF:2021njg}, lattice-field-theory sampling \cite{Kanwar:2020xzo}, and physics-informed studies of differential equations \cite{Raissi:2017zsi,karniadakis2021physics,
	Kou:2025qsg,Kou:2026iau,Dai:2026nzp}. Their flexibility is useful for representing multidimensional observables \cite{Gao:2025dkn,Baihaqi:2025kjd}, but it also introduces an identifiability problem. Because $F_2(x,Q^2)$ is a relatively smooth function over the selected kinematic domain, agreement with the data alone does not demonstrate that a particular physical basis has been uniquely selected, nor does it establish a unique physical interpretation of all fitted quantities. A physics-guided analysis must therefore distinguish the structure imposed by the theoretical basis from the information constrained by the data and examine the resulting description using held-out measurements, simpler data-driven parametrizations, and stability tests.

Motivated by these considerations, we construct a physics-guided phenomenological representation of the selected SLAC $F_2(x,Q^2)$ data. The fermion-resonance component is constrained by the AdS eigenvalue problem and the physical proton-mass condition, whereas the Pomeron-like component provides an effective smooth background. A neural parametrization determines the kinematic mixing $w_\theta(x,Q^2)$, the effective Pomeron normalization $g_{\rm P}(x,Q^2)$, and a regularized log-space correction $\delta_\theta(x,Q^2)$. These quantities control the relative weights and limited deformation of the prescribed holographic components, rather than defining an additional dynamical equation relating the two channels.

A series of training--test, replica, parametrization, kinematic-cut, and data-driven baseline studies indicates a qualitative transition from a Pomeron-like-dominated fitted contribution at lower $x$ to a fermion-resonance-dominated contribution at larger $x$. The location of this transition is definition dependent, with the diagnostics considered here placing it in the approximate region $x\sim0.19$--$0.23$. The fitted value $\alpha_0\simeq1.08$ is compatible with conventional soft-Pomeron phenomenology \cite{Donnachie:1992ny}, although it is not separately identifiable from the flexible Pomeron normalization. Accordingly, the framework is interpreted as a transparent, basis-conditioned phenomenological decomposition of the measured structure function, rather than as a unique determination of an $x$-dependent proton spectrum or a demonstration of predictive superiority over a data-driven parametrization.

The remainder of this paper is organized as follows. Section~\ref{sec:holography} reviews the two holographic contributions and the soft-wall background. Section~\ref{sec:pgnn} defines their phenomenological mixing, the constrained holographic quantities, and the regularized correction. Section~\ref{sec:setup} describes the selected SLAC data, the fixed training--test partition, and the numerical diagnostics. Section~\ref{sec:results} presents the global description of $F_2$, the channel fractions, the effective holographic manifold and excited scales, and the fixed-$\alpha_0$ and profile-$\chi^2$ analyses. The conclusions and limitations of the resulting interpretation are summarized in Section~\ref{sec:conclusion}.

\section{Holographic Basis for a Dual-Channel Description}
\label{sec:holography}

\begin{figure}[htpb]
	\centering
	\includegraphics[width=0.5\textwidth]{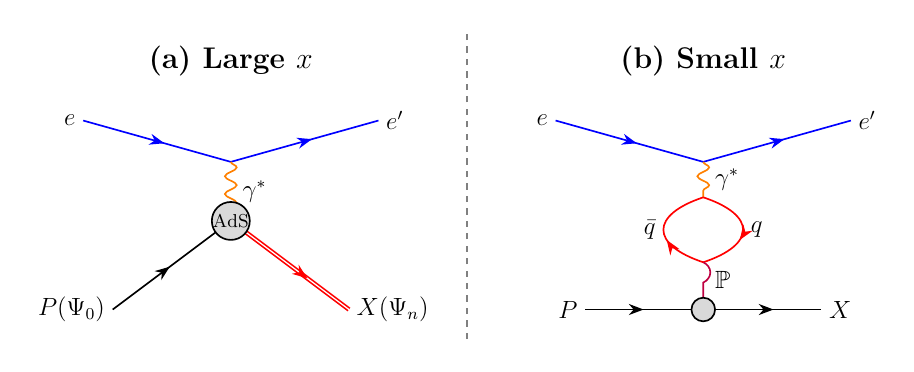}
	\caption{
		Holographic dual-channel picture of deep inelastic scattering.
		(a) In the larger-$x$ regime, the scattering is modeled by an
		$s$-channel fermion-resonance mechanism, in which the virtual photon couples
		to the ground-state bulk fermion field $\Psi_0$ through an AdS overlap
		integral and excites a tower of hadronic resonance states $\Psi_n$ \cite{Braga:2011wa,FolcoCapossoli:2020pks,Gutsche:2017lyu}.
		(b) Toward lower $x$, the smooth background is modeled by a $t$-channel
		holographic Pomeron exchange $\mathbb{P}$, corresponding to closed-string
		diffusion in the AdS bulk \cite{Brower:2006ea,Brower:2010wf}. In the present SLAC kinematic region, this
		contribution should be understood as an effective Pomeron-like
		background rather than as a precise small-$x$ Regge determination.
	}
	\label{fig:feynman}
\end{figure}

This section summarizes the two holographic contributions adopted as prescribed basis functions for the phenomenological description of DIS. The corresponding physical picture is shown schematically in Fig.~\ref{fig:feynman}. The first contribution is associated with $s$-channel transitions between normalizable bulk-fermion states, whereas the second describes a smooth background through $t$-channel holographic Pomeron exchange \cite{Polchinski:2002jw,BallonBayona:2007qr,Braga:2011wa,
	FolcoCapossoli:2020pks,Brower:2006ea,Brower:2010wf}. These two contributions are not derived here as coupled sectors of a new dynamical equation; rather, they provide two existing holographic basis components whose relative contributions are determined phenomenologically in Sec.~\ref{sec:pgnn}. According to the AdS/CFT dictionary \cite{Maldacena:1997re,Gubser:1998bc,Witten:1998qj}, the boundary electromagnetic current and hadronic states are represented by a bulk vector field and normalizable spinor modes in a five-dimensional asymptotically $\text{AdS}_5$ geometry \cite{Polchinski:2002jw,Abidin:2009hr,Gao:2009ze}.

\subsection{Kinematics and the Soft-Wall Background}
\label{subsec:kinematics}

Let $P$ and $q$ denote the proton and virtual-photon four-momenta, respectively, with $P^2 = M_p^2$ and $q^2 = -Q^2 < 0$. The Bjorken variable and the invariant mass of the inclusive hadronic final state are
\begin{equation}
	\begin{aligned}
		x
		&=
		\frac{Q^2}{2P\cdot q},
		\\
		W^2
		&\equiv
		(P+q)^2
		=
		M_p^2
		+
		Q^2\frac{1-x}{x}.
	\end{aligned}
	\label{eq:dis_kinematics}
\end{equation}
The soft-wall construction introduces an infrared scale via the background profile
\begin{equation}
	\Phi(z)=\kappa^2 z^2,
	\label{eq:soft_wall_profile}
\end{equation}
where $z$ is the holographic coordinate and $\kappa^2$ is the dimensionful soft-wall or deformation scale
\cite{Karch:2006pv,Andreev:2006ct,Gutsche:2011vb,
	Brodsky:2014yha}. In the deformed $\text{AdS}_5$
realization used for the fermion sector, this scale enters the effective radial potential and governs the separation of the normalizable hadronic modes
\cite{Braga:2011wa,FolcoCapossoli:2020pks}.

In the underlying holographic construction, $\kappa^2$ and the five-dimensional fermion mass $M_5$ characterize the chosen background and bulk field. The bin-indexed quantities $\kappa_i^2$ and $M_{5,i}$ introduced in the phenomenological analysis should therefore not be interpreted as independent measurements of proton parameters at different values of Bjorken $x$. For $x$ in the $i$th measured bin, they specify an effective point on a constrained family of holographic basis functions, with the ground-state proton mass held fixed.

\subsection{The $s$-channel: Bulk Fermion and Resonances}
\label{subsec:schannel}

At the larger-$x$ end of the selected kinematic region, the first basis component is constructed from transitions of the ground-state bulk fermion to excited normalizable states. The left- and right-chiral radial wave functions satisfy Schr\"odinger-type equations \cite{Gutsche:2011vb,Braga:2011wa,
	FolcoCapossoli:2020pks,Brodsky:2014yha},
\begin{equation}
	-\psi_{L,R}''(z)
	+
	V_{L,R}(z;\kappa^2,M_5)\psi_{L,R}(z)
	=
	M_n^2\psi_{L,R}(z),
	\label{eq:schrodinger}
\end{equation}
where $M_n$ denotes the mass scale associated with the $n$th normalizable
mode. The effective potentials $V_{L,R}$ incorporate the dependence on the
deformed background and the bulk-fermion mass.

For each effective coordinate $\kappa_i^2$, the corresponding value of
$M_{5,i}$ is fixed by the common ground-state condition
\begin{equation}
	M_0(\kappa_i^2,M_{5,i})
	=
	M_p,
	\qquad
	M_p=0.938~\mathrm{GeV}.
	\label{eq:mass_constraint}
\end{equation}
Thus, $M_{5,i}$ is not an independently varying fit parameter. The excited scales $M_{n\geq1}$ obtained at a selected point on this manifold are conditional properties of that effective basis and should not be interpreted as physical resonance pole masses that vary with Bjorken $x$.

Retaining the leading Dirac-type transition current, the relevant left- and right-chiral overlap integrals can be written as \cite{Abidin:2009hr,Gao:2009ze,Gutsche:2017lyu,
	Mamedov:2022tth}
\begin{equation}
	\mathcal{I}_n^{L,R}(Q^2)
	=
	\int dz\,
	\mathcal{J}(Q,z)
	\psi_{L,R,0}(z)
	\psi_{L,R,n}(z),
	\label{eq:fermion_overlap}
\end{equation}
where $\mathcal{J}(Q,z)$ denotes the bulk-to-boundary propagator of the virtual
photon. The dependence of the wave functions on $\kappa_i^2$ and $M_{5,i}$ is
left implicit.

For $x$ in the $i$th bin, the fermion-resonance basis used in the present
analysis is the finite-state expression \cite{Polchinski:2002jw,Braga:2011wa,
	FolcoCapossoli:2020pks}
\begin{equation}
	\begin{aligned}
		F_2^{\rm F}(x,Q^2)
		&=
		\frac{G_{{\rm eff},i}Q^2}
		{8xM_p^2}
		\sum_{n=1}^{11}
		\bigg[
		\left|\mathcal{I}_n^L(Q^2)\right|^2
		+
		\left|\mathcal{I}_n^R(Q^2)\right|^2
		\bigg],
	\end{aligned}
	\label{eq:f2_fermion}
\end{equation}
where the fixed overall normalization factors have been absorbed into the effective fermion-sector normalization $G_{{\rm eff},i}$. The formal spectral derivation of the transition sum is reviewed in Appendix~\ref{app:tensor}.

Equation~\eqref{eq:f2_fermion} retains the eleven excited states $n=1,\ldots,11$. No Gaussian or Breit--Wigner resonance profile is introduced in the central construction. It should therefore be regarded as a phenomenological finite-state basis rather than a first-principles description of physical resonance widths.

The retained current contains the leading Dirac-type transition form factor, while the Pauli-type term and subleading tensor structures are omitted in the large-$Q^2$ reduction \cite{Abidin:2009hr,Gutsche:2017lyu,Mamedov:2022tth}. Because the selected data extend to $Q^2=0.583~\mathrm{GeV}^2$, this reduction is not uniformly controlled over the entire fitted domain. The lowest-$Q^2$ results consequently represent an application of the effective fermion-resonance basis rather than a complete low-$Q^2$ current calculation. The corresponding restricted-$Q^2$ stability diagnostics are discussed in Sec.~\ref{sec:results}.

\subsection{The $t$-channel: Holographic Pomeron Exchange}
\label{subsec:tchannel}

At the lower end of the measured $x$ range, a smooth energy-dependent background is described by holographic Pomeron exchange. In the corresponding bulk picture, the virtual photon generates a bulk electromagnetic profile near the ultraviolet boundary, which interacts with the proton profile via a $t$-channel closed-string or graviton exchange
\cite{Brower:2006ea,Hatta:2007he,BallonBayona:2007rs,
	Brower:2010wf,FolcoCapossoli:2015hub}. This contribution is used here as an effective Pomeron-like basis component; within the present SLAC kinematic regime, it is not interpreted as a precision determination of asymptotic small-$x$ dynamics \cite{Jorrin:2022lua}.

Using the optical theorem, the corresponding holographic integral may be
written, up to an overall normalization convention, as \cite{Brower:2010wf,Watanabe:2019zny,Borsa:2023tqr}
\begin{equation}
	\begin{aligned}
		I_2^{\rm P}(x,Q^2)
		&\propto
		\frac{Q^2}{4\pi^2\alpha_{\rm em}}
		\int dz
		\int dz'\,
		P_{13}(z,Q)
		P_{24}(z',M_p)\\
		&\times
		\operatorname{Im}
		\chi(\hat{s},z,z'),
		\\
		\hat{s}
		&=
		\frac{Q^2}{x},
	\end{aligned}
	\label{eq:f2_pomeron}
\end{equation}
where $P_{13}$ and $P_{24}$ denote the virtual-photon and proton bulk profiles, respectively. The kernel $\chi(\hat{s},z,z')$ describes diffusion in the holographic direction and provides the smooth energy dependence of the exchanged closed-string contribution \cite{Brower:2006ea,Brower:2010wf,Watanabe:2019zny}.

The Pomeron-like contribution entering the phenomenological combination is defined by
\begin{equation}
	\begin{aligned}
		F_2^{\rm P}(x,Q^2)
		&=
		\mathcal{G}_{\rm P}(x,Q^2)
		I_2^{\rm P}(x,Q^2),
		\\
		\mathcal{G}_{\rm P}(x,Q^2)
		&=
		g_P(x,Q^2)
		\left(
		\frac{Q^2}{xQ_0^2}
		\right)^{\alpha_0-1},
		\qquad
		Q_0^2=3~\mathrm{GeV}^2.
	\end{aligned}
	\label{eq:F2p}
\end{equation}
The reference scale $Q_0^2$ renders the base of the noninteger power dimensionless. A change in this fixed reference scale can be absorbed into the effective normalization $g_P(x,Q^2)$ and therefore does not alter the predicted structure function.

Because $g_P(x,Q^2)$ is flexible, variations in $\alpha_0$ can be compensated by corresponding changes in $g_P(x,Q^2)$. The quantity constrained by the present data is therefore the combined factor
$\mathcal{G}_{\rm P}(x,Q^2)$, rather than $g_P(x,Q^2)$ and $\alpha_0$
separately. Accordingly, $\alpha_0$ is treated as an effective parameter
whose compatibility with soft-Pomeron phenomenology can be assessed, not
as an independently determined intercept \cite{Donnachie:1992ny,Jorrin:2022lua}.

Equations~\eqref{eq:f2_fermion} and \eqref{eq:F2p} define the two prescribed
holographic basis contributions used in the phenomenological combination
below. Their labels refer to the holographic mechanisms from which
they are constructed and should not be interpreted as a direct,
model-independent separation of the measured structure function into
valence-quark and sea-quark or gluon distributions.

\section{Physics-Guided Phenomenological Representation}
\label{sec:pgnn}

This section defines the phenomenological representation used to combine the
two holographic basis contributions introduced in
Sec.~\ref{sec:holography}. The fermion-resonance sector provides
$F_2^{\rm F}(x,Q^2)$, while the Pomeron-like sector provides
$I_2^{\rm P}(x,Q^2)$ and the associated contribution
$F_2^{\rm P}(x,Q^2)$ defined in Eq.~\eqref{eq:F2p}. Their relative weighting
and limited deformation are described by a neural parametrization.

\begin{figure*}[htpb]
	\centering
	\includegraphics[width=0.7\linewidth]{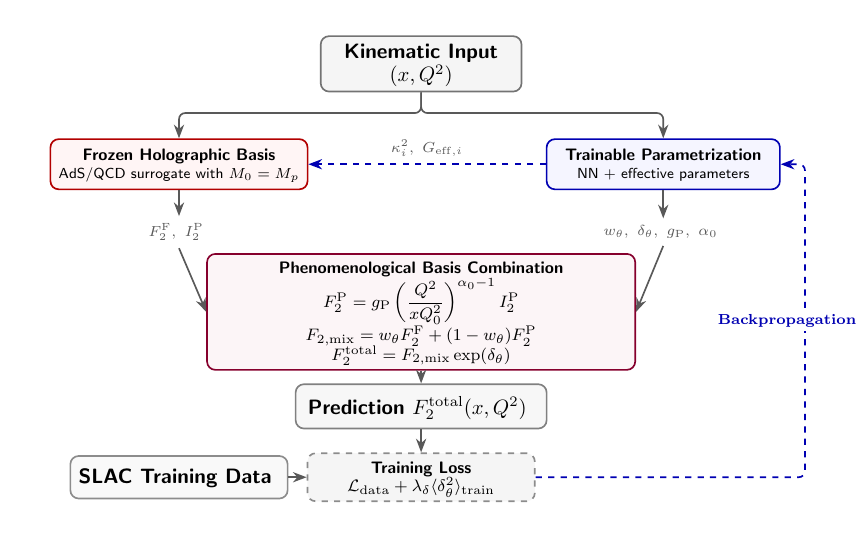}
	\caption{
		Schematic representation of the basis-conditioned, physics-guided phenomenological fit. The exact holographic solver is used offline to generate a theory grid, which is represented during the SLAC fit by the frozen surrogate $\mathcal{S}_{\varphi}$. The data-trained block contains the neural-network outputs $w_\theta$, $g_{\rm P}$, and $\delta_\theta$, the bin-indexed effective quantities $\kappa_i^2$ and $G_{{\rm eff},i}$, and the global effective parameter $\alpha_0$. For each $\kappa_i^2$, the associated bulk mass $M_{5,i}$ is fixed by the physical proton-mass condition $M_0(\kappa_i^2,M_{5,i})=M_p$. The surrogate returns the fermion-resonance basis $F_2^{\rm F}$ and the Pomeron-like basis $I_2^{\rm P}$. The latter is converted into $F_2^{\rm P}$ through the combined factor $\mathcal{G}_{\rm P}$ before the two prescribed basis contributions are combined phenomenologically. Only the training subset enters the optimization objective; the held-out points are used solely for diagnostics. }
	\label{fig:AdS_ML_F2_structure}
\end{figure*}

The physical guidance enters through the prescribed holographic functional
forms and the proton-mass constraint. The neural component does not introduce
a new dynamical relation between the two channels, nor does the data fit solve
the exact AdS eigenvalue problem at each optimization step. Instead, it
provides a kinematic parametrization conditioned on the selected holographic
basis. This use of prescribed physical structure together with data-trained
components belongs to the broader class of theory-guided and hybrid
physics--machine-learning models
\cite{Karpatne_2017,willard2022integrating}. The resulting construction is shown schematically in
Fig.~\ref{fig:AdS_ML_F2_structure}.

\subsection{Kinematic Parametrization and Basis Combination}
\label{subsec:fusion}

The neural parametrization takes as input kinematic features constructed from $x$ and $Q^2$ and returns three functions,
\begin{equation}
	\begin{aligned}
		w_\theta(x,Q^2)
		&\in (0,1),
		\\
		g_P(x,Q^2)
		&>0,
		\\
		\delta_\theta(x,Q^2)
		&\in \mathbb{R}.
	\end{aligned}
	\label{eq:neural_outputs}
\end{equation}
Here, $w_\theta$ is a bounded gate that controls the relative weighting of the two basis components, $g_P$ is the effective normalization entering the Pomeron-like contribution in Eq.~\eqref{eq:F2p}, and $\delta_\theta$ is a multiplicative correction expressed in logarithmic space. The exact input features, network architecture, and output transformations are specified in Sec.~\ref{sec:setup}.

The two holographic contributions are first combined as follows
\begin{equation}
	\begin{aligned}
		F_{2,\rm mix}(x,Q^2)
		&=
		w_\theta(x,Q^2)F_2^{\rm F}(x,Q^2)
		\\
		&\quad+
		\left[1-w_\theta(x,Q^2)\right]
		F_2^{\rm P}(x,Q^2).
	\end{aligned}
	\label{eq:fusion_mix}
\end{equation}
The final prediction is then defined by
\begin{equation}
	F_2^{\rm total}(x,Q^2)
	=
	F_{2,\rm mix}(x,Q^2)
	\exp\!\left[\delta_\theta(x,Q^2)\right].
	\label{eq:fusion_total}
\end{equation}
The exponential form makes $\delta_\theta$ additive in $\log F_2$ and
preserves the positivity of the prediction whenever the mixed basis
contribution is positive.

Equations~\eqref{eq:fusion_mix} and \eqref{eq:fusion_total} constitute a
phenomenological ansatz rather than a consequence of a common five-dimensional
action. In particular, $w_\theta$ is not a microscopic transition probability,
and the weighted combination is not a field-theoretic subtraction theorem for
the possible overlap between the two holographic descriptions. Its role is to
define a smooth, data-constrained interpolation within the selected basis.

\subsection{Constrained Holographic Coordinates and Regularization}
\label{subsec:loss}

For a data point belonging to the measured Bjorken-$x_i$ bin, the fermion-resonance basis is evaluated using the effective coordinate $\kappa_i^2$ and the normalization $G_{{\rm eff},i}$. The associated bulk mass $M_{5,i}$ is not fitted independently; instead, it is fixed by the common ground-state condition in Eq.~\eqref{eq:mass_constraint}. The quantities $\kappa_i^2$ and $M_{5,i}$ therefore specify a point on the proton-mass-constrained holographic manifold. They are not interpreted as fundamental proton parameters that physically vary with Bjorken $x$.

The quantities fitted to the SLAC training data \cite{Whitlow:1990gk,Whitlow:1991uw} may be summarized as
\begin{equation}
	\Theta_{\rm fit}
	=
	\left\{
	\theta_{\rm NN},
	\left\{
	\kappa_i^2,
	G_{{\rm eff},i}
	\right\}_{i=1}^{N_x},
	\alpha_0
	\right\},
	\label{eq:trainable_set}
\end{equation}
where $\theta_{\rm NN}$ denotes the parameters governing
$w_\theta$, $g_P$, and $\delta_\theta$, and $N_x$ is the number of measured
Bjorken-$x$ bins. The effective intercept $\alpha_0$ is varied in the
reference fit and held fixed in the profile-style refits described in
Sec.~\ref{sec:results}, following the general logic of profile-based
identifiability diagnostics \cite{raue2009structural}. Neither $M_{5,i}$ nor the parameters of the frozen
holographic surrogate belongs to $\Theta_{\rm fit}$.

The status of the quantities entering the representation is summarized in
Table~\ref{tab:parameters}. The proton-mass condition removes $M_{5,i}$ as an
independent fit quantity, but it does not, by itself, make the complete
regression low-dimensional because the neural and bin-indexed
parametrizations remain flexible.

\begin{table*}[htpb]
	\centering
	\caption{
		Status and interpretation of the quantities entering the holographic-basis fit to the SLAC data.
	}
	\label{tab:parameters}
	
	\begin{tabular}{
			@{}l
			@{\hspace{0.015\textwidth}}l
			@{\hspace{0.015\textwidth}}l@{}
		}
		\toprule
		
		\parbox[t]{0.18\textwidth}{\raggedright Quantity}
		&
		\parbox[t]{0.20\textwidth}{\raggedright Status in the data fit}
		&
		\parbox[t]{0.54\textwidth}{\raggedright Role and interpretation}
		\\
		
		\midrule
		
		\parbox[t]{0.18\textwidth}{\raggedright
			$F_2^{\rm F}$ and $I_2^{\rm P}$}
		&
		\parbox[t]{0.20\textwidth}{\raggedright
			Prescribed basis functions}
		&
		\parbox[t]{0.54\textwidth}{\raggedright
			Holographic fermion-resonance and Pomeron-like basis components}
		\\
		
		\parbox[t]{0.18\textwidth}{\raggedright
			$\theta_{\rm NN}$}
		&
		\parbox[t]{0.20\textwidth}{\raggedright
			Fitted to the training data}
		&
		\parbox[t]{0.54\textwidth}{\raggedright
			Controls $w_\theta(x,Q^2)$, $g_P(x,Q^2)$, and
			$\delta_\theta(x,Q^2)$}
		\\
		
		\parbox[t]{0.18\textwidth}{\raggedright
			$\kappa_i^2$}
		&
		\parbox[t]{0.20\textwidth}{\raggedright
			Fitted effective coordinate}
		&
		\parbox[t]{0.54\textwidth}{\raggedright
			Selects a point in the constrained holographic basis manifold
			for the $i$th measured $x$ bin}
		\\
		
		\parbox[t]{0.18\textwidth}{\raggedright
			$G_{{\rm eff},i}$}
		&
		\parbox[t]{0.20\textwidth}{\raggedright
			Fitted effective normalization}
		&
		\parbox[t]{0.54\textwidth}{\raggedright
			Controls the normalization of the fermion-resonance basis in
			the $i$th measured $x$ bin}
		\\
		
		\parbox[t]{0.18\textwidth}{\raggedright
			$\alpha_0$}
		&
		\parbox[t]{0.20\textwidth}{\raggedright
			Fitted in the reference analysis}
		&
		\parbox[t]{0.54\textwidth}{\raggedright
			Effective exponent entering the Pomeron-like contribution;
			not separately identifiable from the flexible normalization
			$g_P$}
		\\
		
		\parbox[t]{0.18\textwidth}{\raggedright
			$M_{5,i}$}
		&
		\parbox[t]{0.20\textwidth}{\raggedright
			Constrained}
		&
		\parbox[t]{0.54\textwidth}{\raggedright
			Determined from $\kappa_i^2$ by the physical proton-mass
			condition}
		\\
		
		\parbox[t]{0.18\textwidth}{\raggedright
			$M_p$ and $Q_0^2$}
		&
		\parbox[t]{0.20\textwidth}{\raggedright
			Fixed}
		&
		\parbox[t]{0.54\textwidth}{\raggedright
			Physical proton mass and reference scale used in the
			dimensionless Pomeron factor}
		\\
		
		\bottomrule
	\end{tabular}
\end{table*}

The residual correction is controlled by a quadratic Tikhonov-type penalty \cite{tikhonov1977solutions} in the regularized objective.
\begin{equation}
	\mathcal{L}(\Theta_{\rm fit})
	=
	\mathcal{L}_{\rm data}
	+
	\lambda_\delta
	\left\langle
	\delta_\theta^2
	\right\rangle_{\rm train},
	\label{eq:loss}
\end{equation}
where the average in the second term is taken only over the training points. The exact weighted log-space definition of $\mathcal{L}_{\rm data}$, the central value of $\lambda_\delta$, and the training--test construction are given in Sec.~\ref{sec:setup}.

The penalty in Eq.~\eqref{eq:loss} introduces an explicit trade-off between adherence to the prescribed holographic basis and additional phenomenological flexibility. Consequently, a small fitted $\delta_\theta$ cannot by itself be taken as evidence that the holographic basis is dynamically complete, because its magnitude is also influenced by $\lambda_\delta$. Its interpretation must therefore be accompanied by the residual-penalty scan and the data-only comparison.

\subsection{Basis Fractions and Identifiability}
\label{subsec:fractions}

The gate $w_\theta$ does not generally equal the physical fraction of the mixed basis associated with the fermion-resonance component. The two basis functions can have different magnitudes even when their gate coefficients are equal. We therefore define the normalized mixed-basis fractions by
\begin{equation}
	\begin{aligned}
		C_{\rm F}(x,Q^2)
		&=
		\frac{
			w_\theta(x,Q^2)F_2^{\rm F}(x,Q^2)
		}{
			F_{2,\rm mix}(x,Q^2)
		},
		\\
		C_{\rm P}(x,Q^2)
		&=
		\frac{
			\left[1-w_\theta(x,Q^2)\right]F_2^{\rm P}(x,Q^2)
		}{
			F_{2,\rm mix}(x,Q^2)
		},
		\\
		&C_{\rm F}(x,Q^2)+C_{\rm P}(x,Q^2)
		=
		1.
	\end{aligned}
	\label{eq:channel_fractions}
\end{equation}
Because the same factor $\exp[\delta_\theta(x,Q^2)]$ multiplies both components, it cancels in the normalized fractions. Thus, $C_{\rm F}$ and $C_{\rm P}$ characterize the composition of the prescribed holographic mixture, whereas $\delta_\theta$ characterizes the subsequent common deformation of that mixture.

The relative magnitude of the residual deformation is quantified by
\begin{equation}
	\begin{aligned}
		R_\delta(x,Q^2)
		&=
		\left|
		\frac{
			F_2^{\rm total}(x,Q^2)
			-
			F_{2,\rm mix}(x,Q^2)
		}{
			F_2^{\rm total}(x,Q^2)
		}
		\right|
		\\
		&=
		\left|
		1-
		\exp\!\left[-\delta_\theta(x,Q^2)\right]
		\right|.
	\end{aligned}
	\label{eq:residual_fraction}
\end{equation}

Two distinct crossover criteria can therefore be considered. The gate
criterion is defined by $w_\theta=1/2$, whereas equality of the final
mixed-basis contributions is defined by $C_{\rm F}=C_{\rm P}$. These
conditions need not occur at the same value of $x$, because the latter also
depends on the relative magnitudes of $F_2^{\rm F}$ and $F_2^{\rm P}$.
Neither criterion should be interpreted as defining a universal physical boundary.

A second identifiability limitation follows from Eq.~\eqref{eq:F2p}.
The data constrain the combined factor
$\mathcal{G}_{\rm P}(x,Q^2)$ rather than $\alpha_0$ and
$g_P(x,Q^2)$ separately. More generally, $w_\theta$, $g_P$, and
$\delta_\theta$ all affect the same observable and can therefore partially compensate
for one another. Such compensating directions are characteristic of parameter
non-identifiability in flexible multiparameter models
\cite{Raman:2016oes,raue2009structural}. The decomposition obtained from
Eq.~\eqref{eq:channel_fractions} is therefore conditional on the selected
holographic basis, neural parametrization, and residual penalty. It should
not be interpreted as a model-independent separation of the measured
structure function into underlying partonic components.

\section{Numerical Setup, Data Partition, and Diagnostics}
\label{sec:setup}

This section specifies the experimental sample, the fixed training--held-out partition, the frozen numerical representation of the holographic basis, and the optimization and reporting procedures. The implementation separates the offline evaluation of the holographic basis from the quantities calibrated using the SLAC data. The held-out sample and all robustness variations are used solely for diagnostic purposes and do not determine the reference fit.

\subsection{SLAC Sample and Fixed Held-Out Partition}
\label{subsec:data}

The experimental input consists of unpolarized proton structure-function data, $F_2^p(x,Q^2)$, obtained from the SLAC fixed-target analysis of Whitlow \textit{et al.} \cite{Whitlow:1990gk,Whitlow:1991uw}. For each data point, the invariant hadronic mass is evaluated using Eq.~\eqref{eq:dis_kinematics}. The nominal selection requires
\begin{equation}
	x>10^{-3},
	\qquad
	W^2>1.15~\mathrm{GeV}^2.
	\label{eq:data_selection}
\end{equation}
No additional $Q^2$ cut is imposed on the input table in the reference analysis. After selection, the retained sample contains ten measured Bjorken-$x$ bins and $N_{\rm data}=187$ points, with the realized kinematic coverage
\begin{equation}
	\begin{aligned}
		0.080
		&\leq
		x
		\leq
		0.850,
		\\
		0.583~\mathrm{GeV}^2
		&\leq
		Q^2
		\leq
		29.174~\mathrm{GeV}^2,
		\\
		2.665~\mathrm{GeV}^2
		&\leq
		W^2
		\leq
		28.698~\mathrm{GeV}^2.
	\end{aligned}
	\label{eq:data_range}
\end{equation}
Thus, although the nominal lower cut is
$W^2>1.15~\mathrm{GeV}^2$, the smallest value present in the fitted
sample is
\begin{equation}
	W_{\min}^2
	=
	2.664844~\mathrm{GeV}^2.
	\label{eq:actual_w2_min}
\end{equation}

The quantity denoted by $\sigma_i$ below is the total experimental uncertainty provided for the $i$th point in the input table. The present analysis uses these uncertainties as diagonal pointwise weights. A correlated experimental covariance matrix is not reconstructed; consequently, the reported goodness-of-fit quantities should not be interpreted as arising from a complete correlated likelihood.

A fixed partition is constructed independently within each measured Bjorken-$x$ bin. For a bin containing $N_j$ points, a pseudo-random permutation with seed $42$ is generated and
\begin{equation}
	N_{{\rm test},j}
	=
	\left\lfloor
	0.20N_j
	\right\rfloor
	\label{eq:test_count}
\end{equation}
Points are assigned to the held-out sample. The remaining points are assigned
to training. This procedure yields
\begin{equation}
	N_{\rm train}=153,
	\qquad
	N_{\rm test}=34.
	\label{eq:split_counts}
\end{equation}

The held-out points do not contribute to the optimized loss, learning-rate scheduler, stopping criterion, hyperparameter tuning, or selection of the reference fit. Because every measured $x$ bin is represented in both subsets, the held-out statistic evaluates interpolation in $Q^2$ within already observed $x$ bins. It does not evaluate extrapolation to an unseen Bjorken-$x$ region or to a different experiment.

\subsection{Offline Holographic Calculation and Frozen Surrogate}
\label{subsec:surrogate}

The holographic basis is first evaluated independently of the experimental
fit. For each selected value of $\kappa^2$, the exact eigensolver determines
$M_5$ through the proton-mass constraint in Eq.~\eqref{eq:mass_constraint},
after which the fermion-resonance and Pomeron-like basis quantities are
computed on the theory grid.

The grid contains ten linearly spaced values of $x$, ten logarithmically
spaced values of $Q^2$, eight values of $\kappa^2$, and five values of
$G_{\rm eff}$. Its domain is
\begin{equation}
	\begin{aligned}
		0.01
		&\leq
		x
		\leq
		0.90,
		&
		1~\mathrm{GeV}^2
		&\leq
		Q^2
		\leq
		200~\mathrm{GeV}^2,
		\\
		0.10~\mathrm{GeV}^2
		&\leq
		\kappa^2
		\leq
		0.80~\mathrm{GeV}^2,
		&
		1
		&\leq
		G_{\rm eff}
		\leq
		8.
	\end{aligned}
	\label{eq:surrogate_domain}
\end{equation}
The complete Cartesian grid therefore contains $4000$ theory points.

The repeated solution of the exact eigenvalue problem during the experimental fit is replaced by a fixed numerical representation \cite{kasim2022building}.
\begin{equation}
	\begin{aligned}
		\mathcal{S}_{\varphi}:
		\bigg(
		x,\,
		\ln\frac{Q^2}{1~\mathrm{GeV}^2},\,
		\kappa_i^2,\,
		G_{{\rm eff},i}
		\bigg)
		\longmapsto
		\bigg(
		\ln F_2^{\rm F},\,
		\ln I_2^{\rm P}
		\bigg).
	\end{aligned}
	\label{eq:surrogate_map}
\end{equation}
The surrogate $\mathcal{S}_{\varphi}$ is a fully connected
$4\!-\!64\!-\!128\!-\!2$ network with $\tanh$ activations. It contains
$8898$ parameters and is trained for $500$ epochs on the theory-generated
grid using the Adam optimizer \cite{Kingma:2014vow}, with a learning rate of $5\times10^{-3}$, a batch
size of $256$, and a mean-squared error loss on the two logarithmic outputs. Its
parameters are then frozen before the SLAC data are introduced.

The $8898$ surrogate parameters are therefore not data-trained parameters of
the phenomenological fit. During the subsequent fit, gradients propagate through
the fixed map with respect to its effective inputs $\kappa_i^2$ and
$G_{{\rm eff},i}$, but not through the exact AdS eigensolver. The resulting
workflow is
\begin{equation}
	\begin{aligned}
		\text{exact holographic solver}
		&\longrightarrow
		\text{theory grid}
		\\
		&\longrightarrow
		\text{frozen surrogate}
		\\
		&\longrightarrow
		\text{fit to SLAC data}.
	\end{aligned}
	\label{eq:numerical_workflow}
\end{equation}

The domain in Eq.~\eqref{eq:surrogate_domain} is the surrogate-training domain, not a calibrated emulator uncertainty band \cite{kennedy2001bayesian}. The archived reference implementation did not reserve an independent exact-solver test set from these $4000$ theory points. Consequently, no formal surrogate uncertainty is propagated into the final observables, and the surrogate is not claimed to have uniform precision outside its training domain.

In particular, the $14$ experimental points satisfying $0.583\leq Q^2<1~\mathrm{GeV}^2$ lie below the theory-grid range and therefore require extrapolation in $Q^2$. The restricted-$Q^2$ diagnostics described below are used to test whether the main qualitative conclusions depend on these points, but they do not provide a formal emulator-error estimate.

\subsection{Neural Parametrization and Output Transformations}
\label{subsec:network_config}

The data-trained neural parameterization uses three kinematic inputs.
\begin{equation}
	\mathbf{u}(x,Q^2)
	=
	\left(
	x,\,
	\ln\frac{1}{x},\,
	\ln\frac{Q^2}{Q_0^2}
	\right),
	\qquad
	Q_0^2=3~\mathrm{GeV}^2.
	\label{eq:network_inputs}
\end{equation}
The reference network has architecture $3\!-\!16\!-\!8\!-\!3$ with $\tanh$ activations in the hidden layers. Let $h_\delta$, $h_w$, and $h_g$ denote its three raw outputs. The transformations used in the reference calculation are
\begin{equation}
	\begin{aligned}
		\delta_\theta(x,Q^2)
		&=
		0.8\tanh h_\delta(x,Q^2),
		\\
		w_\theta(x,Q^2)
		&=
		\frac{1}{
			1+
			\exp\left[
			-h_w(x,Q^2)-10(x-0.2)
			\right]
		},
		\\
		\widetilde{g}_P(x,Q^2)
		&=
		\exp\left[
		\operatorname{clip}
		\left(
		h_g(x,Q^2),-2,2
		\right)
		\right].
	\end{aligned}
	\label{eq:output_transformations}
\end{equation}
The fixed reference-scale conversion associated with Eq.~\eqref{eq:F2p} is absorbed through
\begin{equation}
	g_P(x,Q^2)
	=
	\widetilde{g}_P(x,Q^2)
	\left(
	\frac{Q_0^2}{1~\mathrm{GeV}^2}
	\right)^{\alpha_0-1}.
	\label{eq:gp_conversion}
\end{equation}
This rewriting leaves the Pomeron-like contribution numerically unchanged while rendering the noninteger power in Eq.~\eqref{eq:F2p} dimensionless.

Equation~\eqref{eq:output_transformations} makes explicit that the gate is not neutrally parameterized in $x$. The fixed term $10(x-0.2)$ favors an increasing fermion-resonance gate centered at $x=0.2$ before the data-dependent output $h_w$ is added. Therefore, a solution of $w_\theta=1/2$ is conditional on this architectural prior and cannot be interpreted as a transition scale identified independently by the data. The contribution-based fractions $C_{\rm F}$ and $C_{\rm P}$ defined in Eq.~\eqref{eq:channel_fractions} remain distinct from this gate criterion.

For the ten measured $x$ bins, the fit contains ten values of $\kappa_i^2$ and ten values of $G_{{\rm eff},i}$. Their reference initializations are
\begin{equation}
	\kappa_i^2=0.45~\mathrm{GeV}^2,
	\qquad
	G_{{\rm eff},i}=4,
	\qquad
	\alpha_0=1.08.
	\label{eq:parameter_initialization}
\end{equation}
During optimization, the numerical bounds
\begin{equation}
	\begin{aligned}
		0.05~\mathrm{GeV}^2
		&\leq
		\kappa_i^2
		\leq
		1.00~\mathrm{GeV}^2,
		\\
		0.5
		&\leq
		G_{{\rm eff},i}
		\leq
		15,
		\\
		1.01
		&\leq
		\alpha_0
		\leq
		1.35
	\end{aligned}
	\label{eq:parameter_bounds}
\end{equation}
are imposed after each optimization step. These numerical bounds are broader
than the theory-grid ranges in Eq.~\eqref{eq:surrogate_domain}. A surrogate
evaluation outside those ranges is therefore extrapolative and is not
assigned a precision interpretation.

The global neural network contains
\begin{equation}
	\begin{aligned}
		N_{\rm NN}
		&=
		(3\times16+16)
		+
		(16\times8+8)
		+
		(8\times3+3)
		\\
		&=
		227
	\end{aligned}
	\label{eq:nn_parameter_count}
\end{equation}
parameters. Including the twenty bin-indexed quantities and the effective
intercept gives
\begin{equation}
	\begin{aligned}
		N_{\rm fit}
		&=
		227
		+
		20
		+
		1
		\\
		&=
		248
	\end{aligned}
	\label{eq:total_parameter_count}
\end{equation}
Quantities are adjusted using the $153$ training points. The fit is therefore
overparameterized in the conventional regression sense. The proton-mass
constraint and residual regularization restrict its functional behavior, but
they do not justify treating $N_{\rm data}-N_{\rm fit}$ as the conventional
number of statistical degrees of freedom \cite{ye1998measuring}.

\subsection{Optimization Objective and Reporting Statistics}
\label{subsec:objective}

For each training point, the scaled logarithmic residual is
\begin{equation}
	r_i^{\log}
	=
	\frac{
		\ln F_{2,i}^{\rm total}
		-
		\ln F_{2,i}^{\rm exp}
	}{
		s_i
	},
	\qquad
	s_i
	=
	\max\left(
	\frac{\sigma_i}{F_{2,i}^{\rm exp}},
	0.01
	\right).
	\label{eq:log_residual}
\end{equation}
The one-percent floor prevents exceptionally small tabulated relative uncertainties from dominating the optimization. The Huber function used in the fit is \cite{huber1992robust}
\begin{equation}
	\operatorname{Huber}_{\Delta}(r)
	=
	\begin{cases}
		\dfrac{1}{2}r^2,
		&
		|r|\leq\Delta,
		\\[0.5em]
		\Delta
		\left(
		|r|-\dfrac{\Delta}{2}
		\right),
		&
		|r|>\Delta.
	\end{cases}
	\label{eq:huber_definition}
\end{equation}
The explicit realization of the objective introduced in Eq.~\eqref{eq:loss} is
\begin{equation}
	\mathcal{L}
	=
	\frac{1}{N_{\rm train}}
	\sum_{i\in{\rm train}}
	\operatorname{Huber}_{\Delta=2}
	\left(
	r_i^{\log}
	\right)
	+
	\lambda_\delta
	\left\langle
	\delta_\theta^2
	\right\rangle_{\rm train},
	\qquad
	\lambda_\delta=0.5.
	\label{eq:explicit_loss}
\end{equation}

The global-network parameters, $\kappa_i^2$ and
$G_{{\rm eff},i}$, are optimized with Adam using a learning rate of
$5\times10^{-3}$ and a weight decay of $10^{-4}$. The effective intercept
$\alpha_0$ is assigned a separate learning rate of $2\times10^{-4}$ and no
weight decay. A ReduceLROnPlateau scheduler operates on the training objective,
reducing the learning rate by a factor of $0.5$ after $150$ epochs without
improvement, with a minimum learning rate of $10^{-5}$. The total gradient norm
is clipped at $1$ \cite{pascanu2013difficultytrainingrecurrentneural}, and the reference optimization is run for $5000$ epochs.

The Huber objective is not numerically identical to the diagnostic
$\chi^2/N$. After training, the latter is evaluated in linear $F_2$ space as
\begin{equation}
	\frac{\chi^2_{\mathcal{S}}}{N_{\mathcal{S}}}
	=
	\frac{1}{N_{\mathcal{S}}}
	\sum_{i\in\mathcal{S}}
	\left(
	\frac{
		F_{2,i}^{\rm total}
		-
		F_{2,i}^{\rm exp}
	}{
		\sigma_i
	}
	\right)^2,
	\label{eq:reported_chi2}
\end{equation}
where $\mathcal{S}$ denotes the training, held-out, or complete sample. We also report \cite{de_Myttenaere_2016}
\begin{equation}
	\operatorname{MAPE}_{\mathcal{S}}
	=
	\frac{100}{N_{\mathcal{S}}}
	\sum_{i\in\mathcal{S}}
	\left|
	\frac{
		F_{2,i}^{\rm total}
		-
		F_{2,i}^{\rm exp}
	}{
		F_{2,i}^{\rm exp}
	}
	\right|.
	\label{eq:mape}
\end{equation}

The split-resolved diagnostics of the fixed-reference fit are summarized in Table~\ref{tab:split_metrics}. Because of the flexible parametrization and the absence of a full covariance matrix, $\chi^2/N$ is reported only as a pointwise descriptive statistic, rather than as $\chi^2/{\rm d.o.f.}$ or as the basis for a formal goodness-of-fit probability.

\begin{table}[htpb]
	\centering
	\caption{
		Split-resolved pointwise diagnostics for the fixed reference fit. The
held-out sample is not used during optimization or model selection.
	}
	\label{tab:split_metrics}
	\begin{tabular}{lrrrr}
		\toprule
		Sample
		&
		$N$
		&
		$\chi^2$
		&
		$\chi^2/N$
		&
		MAPE
		\\
		\midrule
		Training
		&
		$153$
		&
		$82.86$
		&
		$0.542$
		&
		$1.61\%$
		\\
		Held-out
		&
		$34$
		&
		$67.66$
		&
		$1.990$
		&
		$2.10\%$
		\\
		Complete
		&
		$187$
		&
		$150.52$
		&
		$0.805$
		&
		$1.70\%$
		\\
		\bottomrule
	\end{tabular}
\end{table}

\subsection{Robustness Protocol}
\label{subsec:robustness_protocol}

Several variations are evaluated independently of the reference fit. First,
the free-$\alpha_0$ calculation is repeated for ten parameter
initializations. Architecture dependence is tested using
$8\!-\!4$ $\tanh$, $16\!-\!8$ $\tanh$, $32\!-\!16$ $\tanh$, and
$16\!-\!8$ SiLU hidden layers \cite{elfwing2017sigmoidweightedlinearunitsneural}, with five initializations for each
architecture. The residual penalty is varied over
$\lambda_\delta=0$, $0.1$, $0.5$, $1$, and $5$. Twenty experimental-data
replicas \cite{DelDebbio:2004xtd} are generated through independent Gaussian fluctuations with widths
$\sigma_i$, and thirteen fixed-$\alpha_0$ refits span
$1.02\leq\alpha_0\leq1.14$ \cite{raue2009structural}.

A purely data-driven baseline uses the same three kinematic inputs, the same
$16\!-\!8$ hidden widths, the same fixed data partition, and matched
initializations, but maps directly to $\ln F_2$ without the holographic basis.
This comparison tests predictive interpolation only; it does not assess the
physical interpretation provided by the holographic representation.

Kinematic sensitivity is examined through independent refits with
$W^2>3~\mathrm{GeV}^2$ and $W^2>4~\mathrm{GeV}^2$. In addition, the frozen
reference prediction is reevaluated after imposing
$Q^2\geq1~\mathrm{GeV}^2$ and $Q^2\geq2~\mathrm{GeV}^2$. The latter are
post-fit subset diagnostics rather than new parameter refits. The corresponding retained
training and held-out counts are $143/30$ and $118/21$, respectively.

The spreads produced by different initializations, architectures, data
replicas, penalties, and kinematic cuts quantify numerical and model
sensitivity. They do not constitute a calibrated Bayesian posterior or a frequentist
confidence interval and therefore are not labeled as a formal theoretical
uncertainty band. Their numerical consequences are presented in
Sec.~\ref{sec:results}.

\section{Results and Discussion}
\label{sec:results}

This section presents the reference fit and the numerical variations defined in Sec.~\ref{sec:setup}. The discussion distinguishes among three levels of inference: the description of the measured structure function, the conditional decomposition within the prescribed holographic basis, and the stability of effective fit quantities under changes in initialization, architecture, regularization, and kinematic selection. The held-out sample and all diagnostic refits remain separate from the optimization of the fixed reference result.

\subsection{Reference Fit and Held-Out Performance}
\label{subsec:global_fit}

\begin{figure}[htpb]
	\centering
	\includegraphics[width=0.49\textwidth]{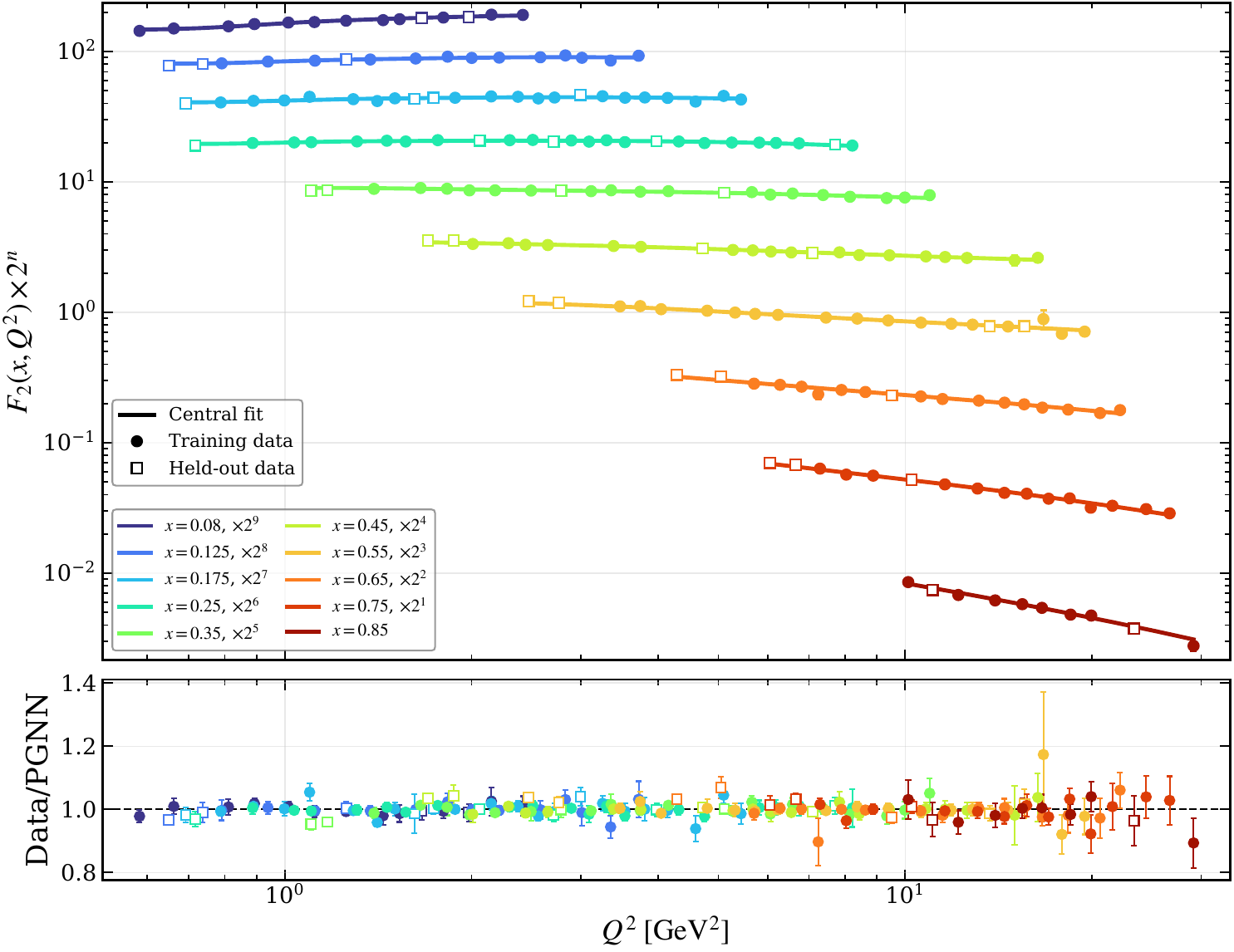}
	\caption{
		Reference holographic-basis fit to the SLAC proton structure-function data \cite{Whitlow:1990gk,Whitlow:1991uw}.
		The upper panel shows $F_2(x,Q^2)$ for the ten measured Bjorken-$x$ bins,
		with successive powers of two applied solely for visual separation. Filled
		circles and open squares denote the training and held-out points, respectively,
		while the solid curves denote the fixed reference prediction. The lower panel
		shows the corresponding data-to-prediction ratios. Experimental
		uncertainties are propagated to both panels. The held-out points are not
		used in the optimization.
	}
	\label{fig:F2_global_fit}
\end{figure}

Figure~\ref{fig:F2_global_fit} compares the reference prediction with the SLAC
measurements in ten Bjorken-$x$ bins \cite{Whitlow:1990gk,Whitlow:1991uw}. The upper panel displays
$F_2(x,Q^2)$ with successive vertical offsets given by powers of two. Filled
circles denote the training points, whereas open squares denote the held-out
points. The lower panel shows the unshifted ratio of each measurement to
the reference prediction.

The split-resolved statistics are presented in Table~\ref{tab:split_metrics}. For the reference fit,
\begin{equation}
	\begin{aligned}
		\frac{\chi^2_{\rm train}}{N_{\rm train}}
		&=
		0.542,
		&
		\operatorname{MAPE}_{\rm train}
		&=
		1.61\%,
		\\
		\frac{\chi^2_{\rm test}}{N_{\rm test}}
		&=
		1.990,
		&
		\operatorname{MAPE}_{\rm test}
		&=
		2.10\%,
		\\
		\frac{\chi^2_{\rm all}}{N_{\rm all}}
		&=
		0.805,
		&
		\operatorname{MAPE}_{\rm all}
		&=
		1.70\%.
	\end{aligned}
	\label{eq:reference_fit_metrics}
\end{equation}
The held-out $\chi^2/N$ is appreciably larger than its training counterpart, as expected for a flexible parametrization with more adjusted quantities than training points. Nevertheless, the held-out MAPE remains close to the training value, and Fig.~\ref{fig:F2_global_fit} shows that the measured $Q^2$ dependence is reproduced across the observed $x$ bins.

The difference between the training and held-out statistics is retained as a generalization diagnostic rather than being absorbed into a single full-sample number \cite{DelDebbio:2004xtd,NNPDF:2017mvq}. Moreover, the spread from the diagnostic refits is not interpreted as a formal confidence interval. Initialization, data-replica, and model variations are instead reported explicitly below.

\subsection{Conditional Channel Decomposition and Residual Sensitivity}
\label{subsec:channel_results}

The channel composition is evaluated using the mixed-basis fractions defined in Eq.~\eqref{eq:channel_fractions}. To avoid evaluating every $x$ bin at a common $Q^2$ value outside its measured support, the reference scale for each bin is chosen as the median of the measured $Q^2$ values in that bin. The resulting path therefore follows the kinematic support of the SLAC sample rather than a fixed-$Q^2$ slice.

\begin{figure}[htpb]
	\centering
	\includegraphics[width=0.49\textwidth]
	{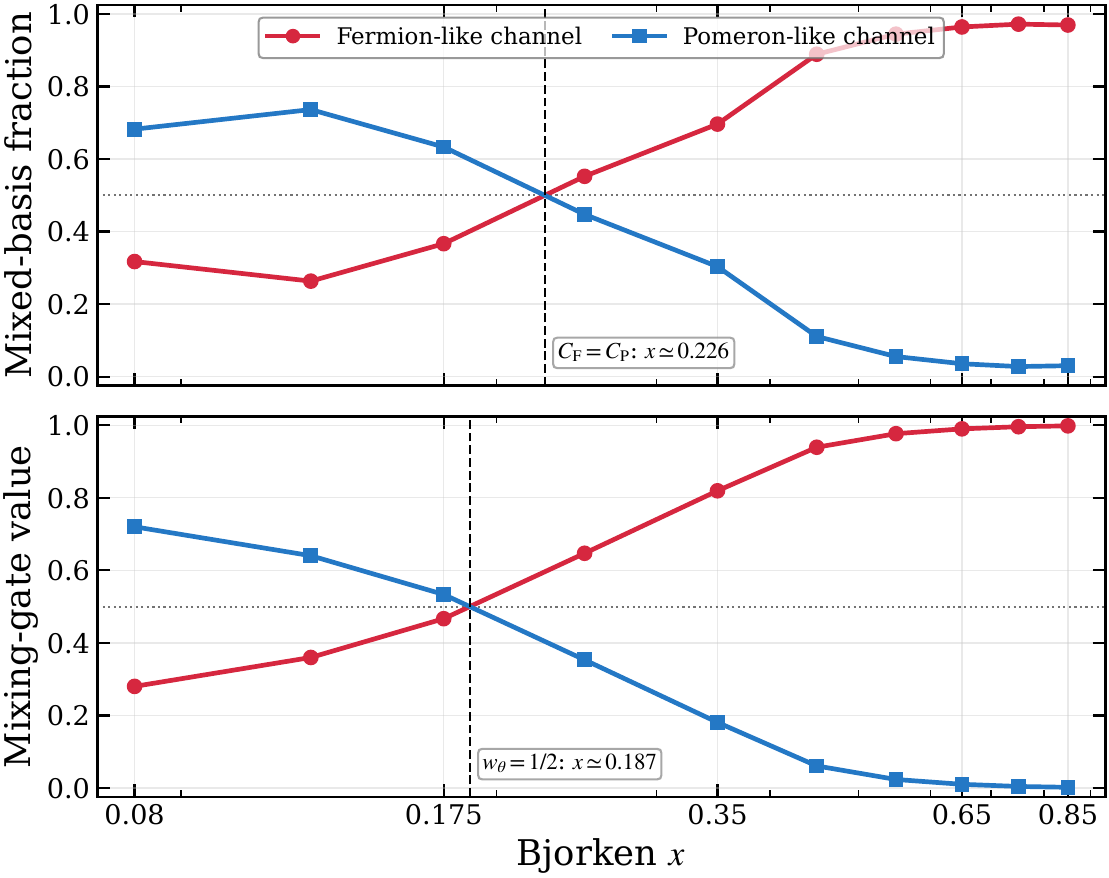}
	\caption{
		Conditional decomposition of the prescribed holographic mixture along the
median measured $Q^2$ in each Bjorken-$x$ bin. The upper panel shows the
mixed-basis fractions $C_{\rm F}$ and $C_{\rm P}$ defined in
Eq.~\eqref{eq:channel_fractions}. Logarithmic interpolation between
neighboring measured $x$ bins yields $C_{\rm F}=C_{\rm P}$ near
$x\simeq0.226$. The lower panel shows the gate $w_\theta$ and its
complement, $1-w_\theta$; the corresponding condition $w_\theta=1/2$
occurs near $x\simeq0.187$. The gate is a diagnostic of the
parametrization and is not itself a contribution fraction.
	}
	\label{fig:channel_decomposition}
\end{figure}

The upper panel of Fig.~\ref{fig:channel_decomposition} shows $C_{\rm F}$ and $C_{\rm P}$ along this bin-median path. At the lowest measured point, $x=0.08$ and $Q_{\rm med}^2=1.3475~\mathrm{GeV}^2$, the Pomeron-like fraction is
\begin{equation}
	C_{\rm P}=0.683,
	\qquad
	C_{\rm F}=0.317.
	\label{eq:low_x_fractions}
\end{equation}
At the largest measured value, $x=0.85$ and $Q_{\rm med}^2=16.032~\mathrm{GeV}^2$, this ordering is reversed:
\begin{equation}
	C_{\rm F}=0.970,
	\qquad
	C_{\rm P}=0.030.
	\label{eq:high_x_fractions}
\end{equation}
The robust qualitative result is therefore a transition from a larger Pomeron-like contribution at the lower end of the measured $x$ range \cite{Brower:2006ea,Brower:2010wf,FolcoCapossoli:2015hub,Watanabe:2019zny} to a fermion-resonance-dominated mixture at large $x$ \cite{Polchinski:2002jw,Braga:2011wa,FolcoCapossoli:2020pks}.

Logarithmic interpolation between neighboring measured $x$ bins yields
\begin{equation}
	\begin{aligned}
		w_\theta=\frac{1}{2}
		&:
		&
		x
		&\simeq
		0.187,
		\\
		C_{\rm F}=C_{\rm P}
		&:
		&
		x
		&\simeq
		0.226.
	\end{aligned}
	\label{eq:two_crossings}
\end{equation}
The separation between these values does not represent a statistical uncertainty associated with a single transition point. Rather, it arises from two different definitions: the first uses only the neural gate, whereas the second also incorporates the relative magnitudes of the two holographic basis functions.

In addition, the gate crossing is influenced by the fixed
$10(x-0.2)$ term in Eq.~\eqref{eq:output_transformations}. It therefore lies
near the center encoded by the selected parametrization and cannot be regarded
as a transition scale independently inferred from the data. The range
$x\sim0.19$--$0.23$ is quoted only as a descriptive, definition-dependent
transition region within the measured sample. Neither endpoint defines a
universal boundary between partonic regimes.

The residual deformation is quantified by
Eq.~\eqref{eq:residual_fraction}. Its magnitude is controlled by
$\lambda_\delta$ and therefore cannot be interpreted independently of the
regularization \cite{tikhonov1977solutions}. Table~\ref{tab:lambda_scan} summarizes the five-initialization
penalty scan.

\begin{table}[htpb]
	\centering
	\caption{
		Median held-out diagnostics from the residual-penalty scan. Each row represents the median across five independent initializations and should not be confused with the single fixed reference fit in Table~\ref{tab:split_metrics}.
	}
	\label{tab:lambda_scan}
	\begin{tabular}{ccc}
		\toprule
		$\lambda_\delta$
		&
		\shortstack{Median held-out\\$\chi^2/N_{\rm test}$}
		&
		\shortstack{Median held-out\\$|\delta_\theta|$}
		\\
		\midrule
		$0$   & $1.741$ & $0.183$ \\
		$0.1$ & $2.018$ & $0.124$ \\
		$0.5$ & $2.143$ & $0.071$ \\
		$1$   & $2.038$ & $0.050$ \\
		$5$   & $1.981$ & $0.020$ \\
		\bottomrule
	\end{tabular}
\end{table}

Increasing $\lambda_\delta$ monotonically suppresses the logarithmic
deformation, whereas the held-out statistic varies only moderately and
non-monotonically. The reference value $\lambda_\delta=0.5$ is therefore not
selected because it minimizes the held-out error. Rather, it defines an
interpretability--flexibility compromise, for which a typical held-out
$|\delta_\theta|$ is approximately $0.071$, corresponding to a multiplicative
deformation of order $7\%$. The unregularized calculation attains a smaller
median held-out statistic only by allowing a substantially larger residual.

A complementary comparison was performed with a purely data-driven network using
the same inputs, hidden widths, data partition, and matched initializations.
Across ten initializations \cite{Karpatne_2017,willard2022integrating,Aarts:2025gyp}, the median training and held-out statistics are
\begin{equation}
	\begin{aligned}
		\text{holographic guided:}
		\qquad
		&
		0.516,
		&
		2.174,
		\\
		\text{data only:}
		\qquad
		&
		0.606,
		&
		1.844,
	\end{aligned}
	\label{eq:data_only_comparison}
\end{equation}
where the two columns denote the training and held-out $\chi^2/N$, respectively.
The data-only parametrization has a slightly smaller median held-out value,
whereas the initialization distributions overlap. Therefore, the present dataset
does not establish a predictive advantage for the holographic basis. Instead,
the demonstrated role of that basis is to provide a conditional
decomposition tied to the two prescribed holographic contributions and the
proton-mass constraint.

\subsection{Proton-Mass Manifold and Effective Excited Scales}
\label{subsec:mass_manifold_results}

\begin{figure*}[htpb]
	\centering
	\includegraphics[width=0.98\textwidth]
	{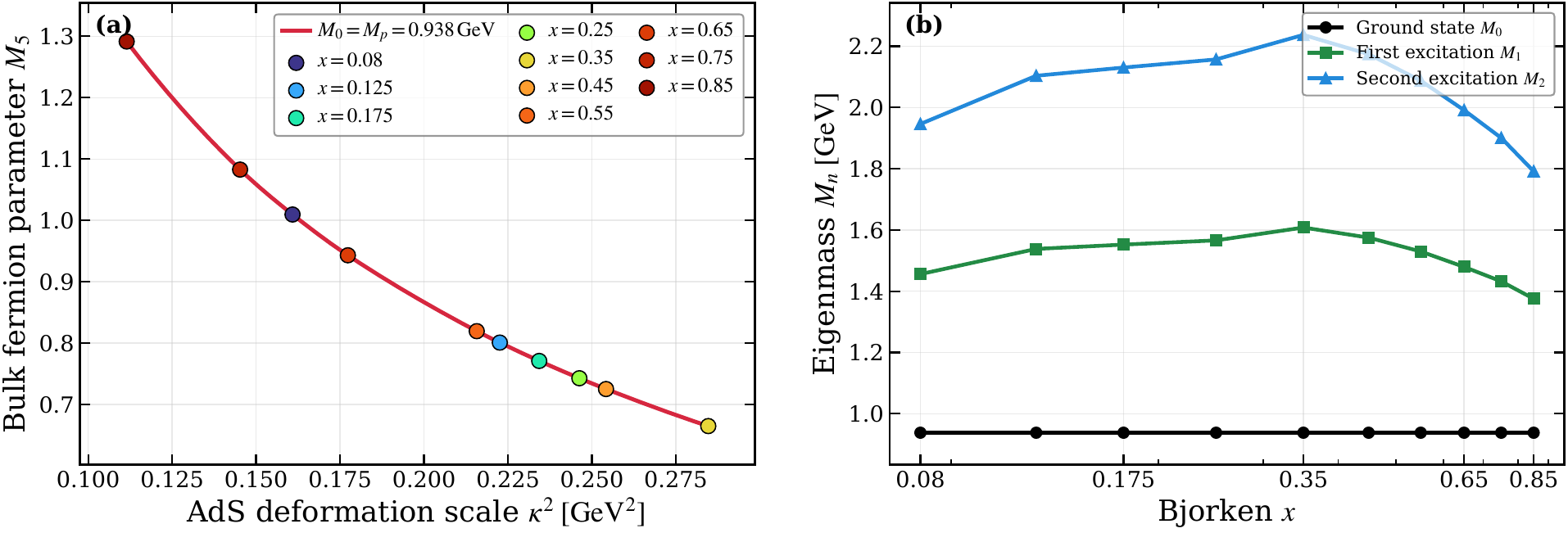}
	\caption{
		Proton-mass-constrained holographic coordinates and their conditional
		outputs. Panel (a) shows the manifold on which
		$M_0=M_p=0.938~\mathrm{GeV}$, together with the effective points
		$(\kappa_i^2,M_{5,i})$ selected for the ten measured Bjorken-$x$ bins.
		Panel (b) shows the ground-state scale and the first two effective excited
		scales obtained at those points. The variation in $M_1$ and $M_2$ is a
		consequence of the bin-indexed coordinates used in the fit; these curves
		are not predictions of physical resonance pole masses or of a literal
		$x$-dependent proton spectrum.
	}
	\label{fig:mass_manifold}
\end{figure*}

The proton-mass condition restricts each effective coordinate pair
$(\kappa_i^2,M_{5,i})$ to the manifold
$M_0(\kappa_i^2,M_{5,i})=M_p$. Figure~\ref{fig:mass_manifold} shows this
curve together with the ten points selected by the measured $x$ bins. Variations in $\kappa_i^2$ therefore move the representation along a common
mass-constrained manifold, rather than altering the physical ground-state
proton mass \cite{Gutsche:2011vb,Braga:2011wa,FolcoCapossoli:2020pks}.

The ground-state value remains fixed by construction, whereas the quantities
$M_1$ and $M_2$ vary along the selected manifold. These higher eigenvalues
provide effective mass scales for the fermion-resonance basis and influence its
$Q^2$ dependence \cite{Gutsche:2017lyu,Mamedov:2022tth}. However, because the underlying $\kappa_i^2$ values are
bin-indexed fit coordinates, the resulting curves are conditional outputs of
the parametrization. They do not constitute independent determinations of the
soft-wall scale, the five-dimensional fermion mass, or the physical excited
proton spectrum.

\subsection{Optimization Stability and Effective Pomeron Intercept}
\label{subsec:alpha_results}

\begin{figure*}[htpb]
	\centering
	\includegraphics[width=\textwidth]
	{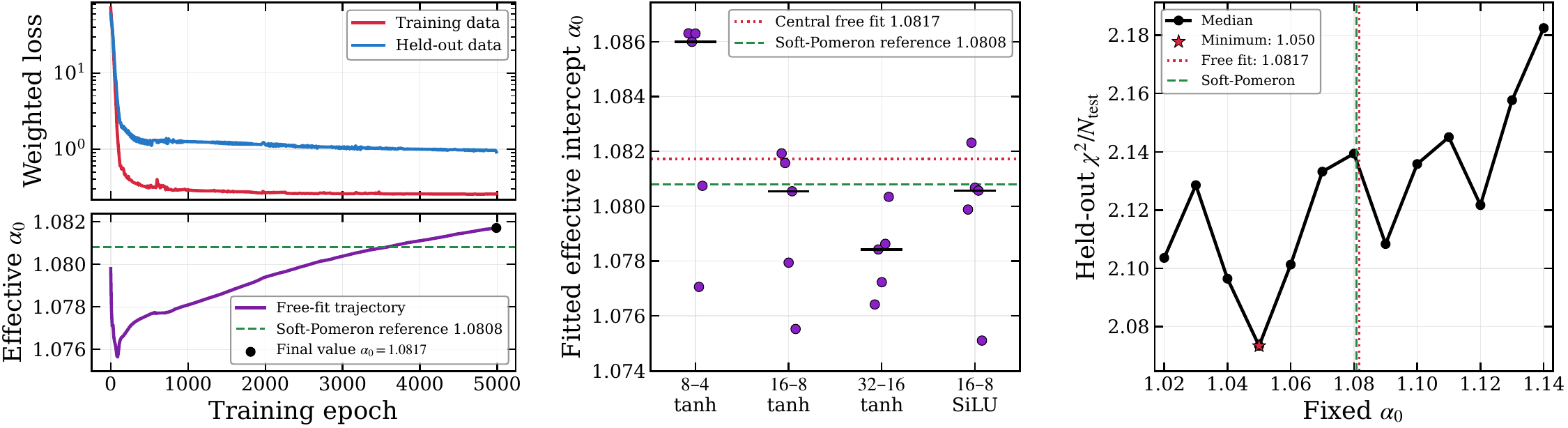}
	\caption{
		Optimization and effective-intercept diagnostics. The upper-left panel
		shows the weighted training and held-out losses during the reference
		optimization, while the lower-left panel shows the free-fit trajectory of
		$\alpha_0$, ending at $\alpha_0=1.0817$. The middle panel shows the fitted
		$\alpha_0$ values obtained from four network architectures and five
		initializations per architecture; horizontal black segments denote the
		corresponding medians. The right panel shows the median held-out
		$\chi^2/N_{\rm test}$ from fixed-$\alpha_0$ refits, with five
		initializations at each scanned value. The sampled minimum occurs at
		$\alpha_0=1.05$. The Donnachie--Landshoff reference,
		$\alpha_{\mathbb{P}}(0)=1.0808$, is shown for comparison. The fixed-parameter
		scan is a robustness diagnostic, not a profile-likelihood confidence
		construction.
	}
	\label{fig:training_alpha0}
\end{figure*}

The convergence and stability diagnostics are presented in
Fig.~\ref{fig:training_alpha0}. The upper-left panel shows the weighted
training and held-out objectives during optimization, while the lower-left
panel shows the trajectory of the global effective intercept. The held-out
curve is evaluated solely for diagnostic purposes and does not feed back into
the optimization.

The free reference fit gives
\begin{equation}
	\alpha_0
	=
	1.0817.
	\label{eq:central_alpha0}
\end{equation}
For comparison, the Donnachie--Landshoff soft-Pomeron reference is $\alpha_{\mathbb{P}}(0)=1.0808$ \cite{Donnachie:1992ny}. The proximity of the two values remains stable under the tested numerical variations. Ten initializations of the reference architecture yield
\begin{equation}
	1.0755
	\leq
	\alpha_0
	\leq
	1.0824,
	\label{eq:alpha_seed_range}
\end{equation}
while the complete scan over four architectures and twenty fits yields
\begin{equation}
	1.0751
	\leq
	\alpha_0
	\leq
	1.0863.
	\label{eq:alpha_architecture_range}
\end{equation}
Twenty replicas of the experimental data \cite{DelDebbio:2004xtd} yield a median
$\alpha_0=1.0804$, with individual values ranging from $1.0716$ to $1.0846$.
These variations support only the limited conclusion that the free fits tend to
remain near the soft-Pomeron reference value.

They do not, however, provide a precision determination of the intercept.
When $\alpha_0$ is fixed between $1.02$ and $1.14$ and all remaining
quantities are reoptimized, the median held-out $\chi^2/N_{\rm test}$ changes
only from $2.073$ to $2.182$. The lowest sampled median occurs at
$\alpha_0=1.05$, rather than at the free-fit endpoint. This near-flatness is
consistent with the identifiability limitation described in
Sec.~\ref{subsec:fractions} \cite{raue2009structural,Raman:2016oes}: the data constrain the combined Pomeron factor
$\mathcal{G}_{\rm P}(x,Q^2)$, but do not separately determine
$\alpha_0$ and $g_P(x,Q^2)$. Consequently, $\alpha_0\simeq1.08$ is a
soft-Pomeron-compatible effective value within this representation, not an
independently extracted intercept \cite{Donnachie:1992ny,Brower:2010wf,Jorrin:2022lua}.

\subsection{Kinematic Stability and Scope}
\label{subsec:kinematic_results}

The effects of the higher-$W^2$ refits and the post-fit $Q^2$ restrictions are summarized in Table~\ref{tab:kinematic_sensitivity}. The first block reports the medians over five independent refits for each $W^2$ selection. The second block reevaluates the fixed reference prediction after removing low-$Q^2$ points and therefore does not involve reoptimization.

\begin{table*}[htpb]
	\centering
	\caption{
		Kinematic-sensitivity diagnostics. Dashes indicate quantities not used in the corresponding comparison. The $W^2$ rows report medians over five refits, whereas the $Q^2$ rows correspond to subsets of the fixed reference prediction with $\alpha_0=1.0817$.
	}
	\label{tab:kinematic_sensitivity}
	\begin{tabular}{llccccc}
		\toprule
		Procedure
		&
		Selection
		&
		$N_{\rm train}$
		&
		$N_{\rm test}$
		&
		$\chi^2_{\rm train}/N_{\rm train}$
		&
		$\chi^2_{\rm test}/N_{\rm test}$
		&
		$\alpha_0$
		\\
		\midrule
		\multicolumn{7}{c}{Independent $W^2$-cut refits}
		\\
		Refit
		&
		$W^2>1.15~\mathrm{GeV}^2$
		&
		$153$
		&
		$34$
		&
		--
		&
		$2.143$
		&
		$1.0805$
		\\
		Refit
		&
		$W^2>3~\mathrm{GeV}^2$
		&
		$152$
		&
		$29$
		&
		--
		&
		$1.355$
		&
		$1.0807$
		\\
		Refit
		&
		$W^2>4~\mathrm{GeV}^2$
		&
		$135$
		&
		$22$
		&
		--
		&
		$0.367$
		&
		$1.0800$
		\\
		\midrule
		\multicolumn{7}{c}{Post-fit subsets of the fixed reference prediction}
		\\
		Fixed prediction
		&
		Full sample
		&
		$153$
		&
		$34$
		&
		$0.542$
		&
		$1.990$
		&
		$1.0817$
		\\
		Fixed prediction
		&
		$Q^2\geq1~\mathrm{GeV}^2$
		&
		$143$
		&
		$30$
		&
		$0.555$
		&
		$1.999$
		&
		$1.0817$
		\\
		Fixed prediction
		&
		$Q^2\geq2~\mathrm{GeV}^2$
		&
		$118$
		&
		$21$
		&
		$0.471$
		&
		$1.531$
		&
		$1.0817$
		\\
		\bottomrule
	\end{tabular}
\end{table*}

The higher-$W^2$ refits leave the median effective intercept near $1.08$, while the held-out statistic improves markedly as the low-$W^2$ points are removed. For the fixed reference prediction, imposing $Q^2\geq1~\mathrm{GeV}^2$ leaves the held-out statistic essentially unchanged, whereas imposing $Q^2\geq2~\mathrm{GeV}^2$ reduces it from $1.990$ to $1.531$. These diagnostics indicate that most of the residual tension is associated with the low-$W^2$ and $Q^2<2~\mathrm{GeV}^2$ portion of the sample. We nevertheless retain these points in the reference analysis to preserve coverage of the low-$Q^2$, low-$W^2$ nonperturbative and transition region.

These observations show that the reported qualitative channel ordering and effective-intercept compatibility are not driven exclusively by the lowest-$W^2$ or lowest-$Q^2$ points. They do not remove the limitations of the leading-tensor fermion basis at $Q^2\simeq0.58~\mathrm{GeV}^2$ \cite{Jaffe:1982pm,Ellis:1982wd,Abidin:2009hr,Gutsche:2017lyu,Mamedov:2022tth}, nor do they provide a formal error estimate for the frozen surrogate below its theory-grid range \cite{kennedy2001bayesian,kasim2022building}.

Taken together, the results support a conditional migration from a Pomeron-like-dominated mixture at the lower end of the measured $x$ range to a fermion-resonance-dominated mixture at large $x$. The data do not determine a unique crossover, do not separately identify $\alpha_0$ and $g_P$, and do not show that the holographic-basis fit predicts the held-out data more accurately than a matched data-only network. These qualifications define the scope within which the holographic decomposition and its effective parameters are interpreted.

\section{Conclusion and Outlook}
\label{sec:conclusion}

We have developed a physics-guided neural representation of the proton
structure function $F_2(x,Q^2)$ that combines two holographically motivated
basis contributions with a trainable mixing gate, an effective
Pomeron-like normalization, and a regularized logarithmic deformation.
The fermionic coordinates are constrained to the proton-mass manifold,
so that the ground-state mass remains fixed at
$M_0=M_p=0.938~\mathrm{GeV}$, while the effective coordinates vary across
the measured Bjorken-$x$ bins. This construction provides a structured
phenomenological representation of the data rather than a
first-principles extraction of unique holographic parameters.

For the fixed stratified partition of the $187$ SLAC points, the reference
fit gives
$\chi^2_{\rm train}/N_{\rm train}=0.542$,
$\chi^2_{\rm test}/N_{\rm test}=1.990$, and
$\chi^2_{\rm all}/N_{\rm all}=0.805$, with corresponding mean absolute
percentage errors of $1.61\%$, $2.10\%$, and $1.70\%$. The held-out points
do not participate in optimization or model selection. Their performance
shows that the fitted representation interpolates the measured $Q^2$
dependence within the ten observed $x$ bins. It does not constitute a test
of extrapolation to an unseen Bjorken-$x$ interval or to an independent
experiment. Moreover, because the fit contains $248$ adjusted quantities
for $153$ training points, the reported $\chi^2/N$ values are descriptive
pointwise statistics and must not be interpreted as conventional
$\chi^2/{\rm d.o.f.}$ values.

Within the prescribed holographic decomposition, the fitted mixture exhibits
a clear qualitative migration across the measured kinematic range. Along the
median measured $Q^2$ in each $x$ bin, the Pomeron-like fraction is
$C_{\rm P}=0.683$ at $x=0.08$, whereas the fermion-resonance fraction reaches
$C_{\rm F}=0.970$ at $x=0.85$. The contribution balance
$C_{\rm F}=C_{\rm P}$ occurs near $x\simeq0.226$, while the neural-gate
condition $w_\theta=1/2$ occurs near $x\simeq0.187$. These two values
represent different diagnostics and do not define the uncertainty interval
of a single transition point. In particular, the gate midpoint is influenced
by the fixed $10(x-0.2)$ term in its parametrization. The data therefore
support a conditional change in channel ordering, but not a unique or
model-independent boundary between dynamical regimes.

The free reference fit gives an effective intercept
$\alpha_0=1.0817$, close to the conventional soft-Pomeron reference value \cite{Donnachie:1992ny}.
The fitted value remains near $1.08$ under the tested changes in
initialization, network architecture, experimental replicas, and kinematic
selection. This numerical stability should not be confused with statistical
identifiability. When $\alpha_0$ is fixed over
$1.02\leq\alpha_0\leq1.14$ and the remaining quantities are reoptimized,
the held-out statistic depends only weakly on the selected value and reaches
its lowest sampled median at $\alpha_0=1.05$. Because the trainable function
$g_P(x,Q^2)$ can compensate for changes in the intercept, the present data
primarily constrain their combined Pomeron-like factor. Thus,
$\alpha_0\simeq1.08$ is compatible with the fitted representation but is not
an independently or precisely extracted Pomeron intercept.

The effective excited scales obtained along the proton-mass manifold must be
interpreted with the same qualification. They control the kinematic behavior
of the fermion-resonance basis, but their $x$ dependence originates from the
bin-indexed effective coordinates used in the fit. They are not predictions
of physical resonance pole masses or evidence for a literal
Bjorken-$x$-dependent proton spectrum. Likewise, the magnitude of the neural
deformation is controlled by the penalty $\lambda_\delta$. Its suppression
under increasing regularization demonstrates that the residual can be
controlled, but does not assign it a model-independent physical meaning.

The comparison with the matched data-only network further limits the scope of
the conclusion. Across the tested initializations, the data-only model gives a
median held-out $\chi^2/N_{\rm test}=1.844$, compared with $2.174$ for the
holographically guided representation. The two initialization distributions
overlap, but the present sample does not demonstrate a predictive advantage
for the holographic basis. Its established role is instead to organize the
fit into physically motivated, mass-constrained components whose conditional
contributions can be examined separately.

Several limitations remain. The available experimental uncertainties are
treated as pointwise total errors because a complete correlated covariance
matrix is not included. The frozen surrogate was trained on a finite
$4000$-point theory grid, but no independent exact-solver validation sample
was reserved, and no emulator uncertainty is propagated to the observables.
Fourteen experimental points lie below its nominal
$Q^2\geq1~\mathrm{GeV}^2$ theory-grid range, and the allowed effective
parameter bounds also extend beyond the surrogate-training domain. In
addition, the leading-tensor reduction used for the fermionic contribution is
not uniformly controlled at the lowest measured values of $Q^2$. The
kinematic-cut studies show that the main qualitative trends are not generated
exclusively by these points, but they do not remove the underlying theoretical
or numerical limitations. The spreads obtained from the performed variations
should therefore be understood as sensitivity diagnostics rather than formal
confidence or credible intervals.

The most immediate numerical improvement is to validate the surrogate on a
new sample generated by the exact holographic solver and excluded from its
training \cite{kennedy2001bayesian,kasim2022building}. The logarithmic root-mean-square error, median relative error, and
$95$th-percentile relative error should be reported separately for both
surrogate outputs. Constraining the phenomenological fit to the validated
interpolation domain, or propagating the resulting emulator error when that
domain is exceeded, would then make the numerical accuracy of the holographic
basis quantitatively assessable.

Further work should also test which conclusions genuinely require the
holographic structure. Useful diagnostics include replacing the two
holographic basis functions with trivial reference functions, removing one
channel at a time, restricting the freedom of $w_\theta$ and $g_P$, eliminating
the fixed gate-center term, and leaving out complete $x$ bins rather than
individual points. These tests would separate information carried by the
holographic basis from that learned by the flexible neural functions.
Replacing the independent bin-indexed coordinates with a smoother or
hierarchical parametrization would additionally reduce the present freedom
and permit more direct tests of parameter identifiability \cite{raue2009structural,Raman:2016oes}.

Finally, incorporating correlated experimental systematics \cite{H1:2015ubc}, extending the
fermionic calculation beyond the leading large-$Q^2$ tensor structure \cite{Abidin:2009hr,Gutsche:2017lyu,Mamedov:2022tth}, and
testing the representation against independent DIS datasets or observables \cite{Dudek:2012vr,AbdulKhalek:2021gbh,
	Anderle:2021wcy,LHeC:2020van}
would be necessary before assigning universal physical significance to the
channel crossover or the effective holographic parameters. At the present
stage, the central result is deliberately narrower: the holographic basis
provides a consistent and interpretable coordinate system for describing the
SLAC structure-function data, within which a stable qualitative change in
channel ordering and a soft-Pomeron-compatible effective intercept are
observed. Both conclusions remain conditional on the selected basis,
parametrization, and kinematic domain.

\begin{acknowledgments}
We are grateful to Dr. M. A. Martín Contreras for fruitful discussions on solving structure functions. This work has been supported by the National Natural Science Foundation of China (Grant No. 12547118), the Research Program of State Key Laboratory of Heavy Ion Science and Technology, Institute of Modern Physics, Chinese Academy of Sciences (Grant No. HIST2025CS08), and the National Key R$\&$D Program of China (Grant No. 2024YFE0109800 and 2024YFE0109802).
\end{acknowledgments}
\appendix

\section{Derivation of the Hadronic Tensor and Structure Functions}
\label{app:tensor}

In this appendix, we derive the leading-tensor spectral representation of the
fermion-like contribution and specify the finite-state prescription used in the
reference calculation. The derivation follows the holographic DIS formalisms of
Refs.~\cite{Polchinski:2002jw,FolcoCapossoli:2020pks}. The Pomeron-like
$t$-channel contribution is defined separately in
Sec.~\ref{subsec:tchannel} and is therefore not repeated here.

The inclusive deep-inelastic-scattering cross section is governed by the
hadronic tensor
\begin{equation}
	W^{\mu\nu}
	=
	\frac{1}{4\pi}
	\int d^4y\,
	e^{iq\cdot y}
	\frac{1}{2}
	\sum_S
	\langle P,S|
	[J^\mu(y),J^\nu(0)]
	|P,S\rangle ,
	\label{eq:app_hadronic_tensor}
\end{equation}
where the factor $1/2$ accounts for the average over the initial proton spin. Lorentz covariance, parity invariance, and electromagnetic current conservation, $q_\mu W^{\mu\nu}=0$, allow the unpolarized tensor to be decomposed as
\begin{equation}
	W^{\mu\nu}
	=
	F_1(x,Q^2)P_1^{\mu\nu}
	+
	\frac{F_2(x,Q^2)}{P\cdot q}
	P_2^{\mu\nu},
	\label{eq:W_decomp}
\end{equation}
where
\begin{equation}
	P_1^{\mu\nu}
	=
	\eta^{\mu\nu}
	-
	\frac{q^\mu q^\nu}{q^2},
	\label{eq:proj1}
\end{equation}
and
\begin{equation}
	P_2^{\mu\nu}
	=
	\left(
	P^\mu
	-
	\frac{P\cdot q}{q^2}q^\mu
	\right)
	\left(
	P^\nu
	-
	\frac{P\cdot q}{q^2}q^\nu
	\right).
	\label{eq:proj2}
\end{equation}
The kinematic variables satisfy
\begin{equation}
	q^2=-Q^2,
	\qquad
	x=\frac{Q^2}{2P\cdot q},
	\qquad
	s=(P+q)^2
	=
	M_p^2+\frac{Q^2(1-x)}{x}.
	\label{eq:app_kinematics}
\end{equation}

To evaluate the $s$-channel contribution, a complete set of holographic fermion states $|P_n,S_n\rangle$ is inserted between the two electromagnetic currents. With the standard relativistic normalization, the corresponding spectral representation is
\begin{widetext}
	\begin{equation}
		W^{\mu\nu}
		=
		\frac{1}{2}
		\sum_{n,S,S_n}
		(2\pi)^4
		\delta^{(4)}(P+q-P_n)
		\langle P,S|J^\mu(0)|P_n,S_n\rangle
		\langle P_n,S_n|J^\nu(0)|P,S\rangle .
		\label{eq:W_sum}
	\end{equation}
\end{widetext}
Here, $M_n^2 = P_n^2$ denotes an eigenvalue of the bulk fermion problem for a fixed choice of holographic parameters.

The matrix element of the electromagnetic current is generated by the interaction between the bulk vector field $A_M(x,z)$, dual to the boundary current, and the initial and excited bulk spinor fields. The minimal interaction term can be written as
\begin{equation}
	S_{\mathrm{int}}
	=
	g_V
	\int d^5x\,
	\sqrt{-g}\,
	e^{-\Phi(z)}
	\bar{\Psi}_n
	e_A^{\,M}\Gamma^A
	A_M
	\Psi_0 ,
	\label{eq:app_interaction}
\end{equation}
where $g_V$ is an effective coupling, $e_A^{\,M}$ is the vielbein, and
$\Gamma^A$ denotes the five-dimensional Dirac matrices.

A current-conserving decomposition of the four-dimensional transition vertex
is
\begin{equation}
	\begin{aligned}
		&\langle P_n,S_n|J^\mu(0)|P,S\rangle
		=
		\bar{u}(P_n,S_n)
		\Gamma_n^\mu(q)
		u(P,S),
		\\
		&\Gamma_n^\mu(q)
		=
		\left(
		\gamma^\mu
		-
		\frac{q^\mu\slashed{q}}{q^2}
		\right)
		f_{1,n}(Q^2)
		+
		\frac{i\sigma^{\mu\nu}q_\nu}{2M_p}
		f_{2,n}(Q^2).
	\end{aligned}
	\label{eq:current_matrix}
\end{equation}
The longitudinal completion in the first term guarantees
$q_\mu\Gamma_n^\mu=0$ and does not contribute when the vertex is contracted
with the transverse projectors in Eqs.~\eqref{eq:proj1} and
\eqref{eq:proj2}.

The Dirac-type transition form factor is determined by the overlap of
the bulk-to-boundary propagator $\mathcal{J}(Q,z)$ with the normalizable
fermion modes. Before transforming to Schrödinger-normalized wave functions,
it is given by
\begin{equation}
	\begin{aligned}
		f_{1,n}(Q^2)
		&=
		\int dz\,
		\sqrt{-g}\,
		e^{-\Phi(z)}
		z\,\mathcal{J}(Q,z)
		\\
		&\quad\times
		\left[
		\psi_{L,0}(z)\psi_{L,n}(z)
		+
		\psi_{R,0}(z)\psi_{R,n}(z)
		\right].
	\end{aligned}
	\label{eq:f1_integral}
\end{equation}
The corresponding normalization factors can be absorbed into the effective
fermion-channel coupling used in the phenomenological basis.

The reference calculation retains only the leading Dirac-type tensor
structure and neglects the Pauli-type transition form factor
$f_{2,n}(Q^2)$. This simplification is motivated by the large-$Q^2$ limit,
in which the Pauli-type term is subleading. However, it is not uniformly
controlled at the lowest momentum transfers in the fitted sample, which
extends to $Q^2=0.583~\mathrm{GeV}^2$. The post-fit diagnostics with
$Q^2\geq1~\mathrm{GeV}^2$ and $Q^2\geq2~\mathrm{GeV}^2$, reported in
Table~\ref{tab:kinematic_sensitivity}, show that the principal qualitative
results are not driven exclusively by the lowest-$Q^2$ points. However,
such stability under these cuts does not provide a quantitative estimate of
the omitted Pauli-type contribution.

After retaining the Dirac-type term, the spin summation can be written as a
Dirac trace:
\begin{widetext}
	\begin{equation}
		\begin{aligned}
			W^{\mu\nu}
			&=
			\frac{1}{2}
			\sum_n
			(2\pi)^4
			\delta^{(4)}(P+q-P_n)
			\left|f_{1,n}(Q^2)\right|^2
			\\
			&\quad\times
			\operatorname{Tr}
			\left[
			(\slashed{P}_n+M_n)
			\gamma^\mu
			(\slashed{P}+M_p)
			\gamma^\nu
			\right]
			+
			\text{longitudinal terms}.
		\end{aligned}
		\label{eq:trace}
	\end{equation}
\end{widetext}
The terms proportional to $q^\mu$ or $q^\nu$ vanish under the transverse projection. The remaining trace is
\begin{equation}
	\begin{aligned}
		\operatorname{Tr}
		\left[
		(\slashed{P}_n+M_n)
		\gamma^\mu
		(\slashed{P}+M_p)
		\gamma^\nu
		\right]
		&=
		4\Big[
		P_n^\mu P^\nu
		+
		P_n^\nu P^\mu
		\\
		&-
		\eta^{\mu\nu}
		(P_n\cdot P-M_nM_p)
		\Big].
	\end{aligned}
	\label{eq:app_trace_result}
\end{equation}

Using $P_n = P + q$ and matching the transverse tensor structures to Eq.~\eqref{eq:W_decomp} yields the formal spectral form
\begin{equation}
	\begin{aligned}
		F_{2,\mathrm{spec}}^{\mathrm{F}}(x,Q^2)
		&=
		\mathcal{C}Q^2
		\sum_n
		\delta(s-M_n^2)
		\mathcal{H}_n(Q^2),
		\\
		\mathcal{H}_n(Q^2)
		&=
		\left|\mathcal{I}_{L,n}(Q^2)\right|^2
		+
		\left|\mathcal{I}_{R,n}(Q^2)\right|^2,
	\end{aligned}
	\label{eq:app_F2_spectral}
\end{equation}
where $\mathcal{C}$ contains the remaining normalization and phase-space
factors. The chiral overlap integrals are defined, in the
Schrödinger-normalized convention used in the numerical calculations, as
\begin{equation}
	\begin{aligned}
			&\mathcal{I}_{\chi,n}
		\left(
		Q^2;\kappa_i^2,M_{5,i}
		\right)
		\\&=
		\int dz\,
		\mathcal{J}
		\left(
		Q,z;\kappa_i^2
		\right)
		\psi_{\chi,0}
		\left(
		z;\kappa_i^2,M_{5,i}
		\right)
		\psi_{\chi,n}
		\left(
		z;\kappa_i^2,M_{5,i}
		\right),
	\end{aligned}
	\label{eq:app_chiral_overlap}
\end{equation}
with $\chi=L,R$. The factors associated with the original curved-space measure are incorporated into the definition and normalization of these Schrödinger modes.

For completeness, in an asymptotic regime in which the excited states become densely spaced, a discrete spectral sum can formally be approximated by
\begin{equation}
	\begin{aligned}
		\sum_n
		\delta(s-M_n^2)
		\mathcal{H}_n
		&\longrightarrow
		\int dn\,
		\delta(s-M_n^2)
		\mathcal{H}_n
		\\
		&=
		\left.
		\frac{\partial n}{\partial M_n^2}
		\mathcal{H}_n
		\right|_{M_n^2=s}.
	\end{aligned}
	\label{eq:continuum_limit}
\end{equation}
For an asymptotically linear trajectory,
$M_n^2 \simeq an + b$, the density of states approaches
$\partial n/\partial M_n^2 \simeq 1/a$. This continuum relation explains the
origin of the density-of-states expression frequently used in analytic
holographic DIS calculations. It is not the prescription used in the
numerical fit of the reference.

Instead, the fermion-like holographic basis employed in this work is defined
directly through the finite overlap sum
\begin{widetext}
\begin{equation}
	\begin{aligned}
		F_2^{\mathrm{F}}
		\left(
		x,Q^2;\kappa_i^2,M_{5,i},G_{{\rm eff},i}
		\right)
		=
		\frac{
			\alpha_{\mathrm{F}}
			G_{{\rm eff},i}
			Q^2
		}{
			8xM_p^2
		}
		\sum_{n=1}^{11}
		\Bigg[
		\left|
		\mathcal{I}_{L,n}
		\left(
		Q^2;\kappa_i^2,M_{5,i}
		\right)
		\right|^2
		+
		\left|
		\mathcal{I}_{R,n}
		\left(
		Q^2;\kappa_i^2,M_{5,i}
		\right)
		\right|^2
		\Bigg],
	\end{aligned}
	\label{eq:app_F2_finite}
\end{equation}
\end{widetext}
where $\alpha_{\mathrm{F}}=1$ is fixed in the reference implementation, and
$G_{{\rm eff},i}$ is the effective fermion-channel normalization associated
with the measured $x_i$ bin. The ground state, $n=0$, represents the incoming
proton and is not included in the excited-state sum.

Equation~\eqref{eq:app_F2_finite} contains eleven discrete excited states,
$n=1,\ldots,11$. Neither a Gaussian replacement nor a Breit--Wigner
replacement of $\delta(s-M_n^2)$ is introduced in the numerical
implementation, and no resonance-smearing width enters the reference
calculation. The finite overlap sum should therefore be understood as a
phenomenological discrete-state basis rather than as a first-principles
calculation of physical resonance widths or line shapes.

For every fixed pair $(\kappa_i^2,M_{5,i})$, the quantities $M_n$ are
eigenvalues of the corresponding holographic problem. In the global fit,
however, $\kappa_i^2$ is an effective parameter associated with the measured
$x_i$ bin, while $M_{5,i}$ is fixed by
\begin{equation}
	M_0(\kappa_i^2,M_{5,i})=M_p.
	\label{eq:app_mass_constraint}
\end{equation}
Consequently, the fitted values denoted by $M_n(x_i)$ are effective holographic scales evaluated at different points on the proton-mass manifold. They should not be interpreted as Bjorken-$x$-dependent physical resonance pole masses. By construction, only the ground-state mass is fixed to the physical proton mass.

The formal spectral representation in Eq.~\eqref{eq:app_F2_spectral} establishes the hadronic-tensor origin of the fermion tower, whereas Eq.~\eqref{eq:app_F2_finite} provides the operational definition of the fermion-like basis used in the reference fit.

\bibliography{refs}

\end{document}